\documentclass[a4paper,10pt]{article}

\usepackage{amssymb}
\usepackage{amsmath}
\usepackage[usenames,dvipsnames]{color}
\usepackage{hyperref}
\usepackage[left=1in,right=1in,top=1in,bottom=1in]{geometry}                  

\hypersetup{
    colorlinks=true,         
    linkcolor=blue,          
    citecolor=red,        
    urlcolor=Violet             
}

\newtheorem{defi}{Definition}
\newtheorem{theo}{Theorem}
\newtheorem{lem}{Lemma}

\newcommand{\mean}[1]{\ensuremath{\lf\langle #1 \rt\rangle }}
\newcommand{\diby}[2]{\ensuremath{\frac{\delta #1}{\delta #2}}}
\newtheorem{prop}{Proposition}
\newcommand{\order}[1]{\ensuremath{\mathcal{O}(#1)}}

\def\be{\begin{equation}}
\def\ee{\end{equation}}
\def\bea{\begin{eqnarray}}
\def\eea{\end{eqnarray}}

\def\lf {\ensuremath{\left}}
\def\rt {\ensuremath{\right}}

\title{Symmetry Doubling: Doubly General Relativity}
\author{\bf Henrique Gomes\footnote{\href{mailto:gomes.ha@gmail.com}{gomes.ha@gmail.com}}\\\it University of California at Davis\\ \it One Shields Avenue Davis, CA, 95616, USA \bigskip\\ \bf Tim Koslowski\footnote{\href{mailto:t.a.koslowski@gmail.com}{tkoslowski@perimeterinstitute.ca}}
\\\it Perimeter Institute for Theoretical Physics\\\it 31 Caroline Street, Waterloo, Ontario N2L 2Y5, Canada}

\begin{document}

\maketitle

\begin{abstract}
  We show that the equivalence of General Relativity and Shape Dynamics can be extended to a theory, that respects the BRST-symmetries of General Relativity as well as the ones of an extended version of Shape Dynamics. This version of Shape Dynamics implements local spatial conformal transformations as well as a local an abstract analogue of special conformal transformations. Standard effective field theory arguments suggest that the definition of a gravity theory should implement this duality between General Relativity and Shape Dynamics, thus the name ``Doubly General Relativity.'' We briefly discuss several consequences: bulk/bulk- duality in classical gravity, experimental falsification of Doubly General Relativity and possible implications for the renormalization of quantum gravity in the effective field theory framework.
\end{abstract}


\section{Introduction}

Two different stories about the same entity are hardly ever consistent with one another. The purpose of this paper  is finding the common truth behind two seemingly inconsistent, and yet internally coherent, paradigms about the nature of classical gravity. The first is the far better known story of a curved spacetime: the simple, plain description of gravity through the Einstein-Hilbert action.  The second is an equally interesting story  about spatial relationalism. 

Spacetime ontology arose historically as Minkowski's intuitive explanation for a hidden symmetry found by Poincar\'{e} in Maxwell's equations. This intuitive geometric picture is the foundation of special relativity and the further development of spacetime geometric concepts was key in the discovery of General Relativity (GR), which after almost a century is still our best account of gravitational phenomena. This spacetime geometric explanation of gravity was almost incomprehensible for generations whose intuitions had been shaped by the success of a Galileian ontology. Yet, the theory is strikingly intuitive and simple, because it can be neatly expressed in terms of pseudo-Riemannian geometry: gravity is curvature of a Lorentzian manifold. Even more appealing is that this is the consequence of Einstein's insight that a point-like observer can not distinguish between an accelerated spacetime frame and a gravitational force. Combined with the global relativity of simultaneity of Special Relativity, this  provides a central feature of GR: all three-dimensional spacelike surfaces are equally good notions of simultaneity, in the sense that physical predictions do not depend on the choice of simultaneity.

A completely different ontology for gravitational phenomena was developed by Barbour upon a reflection of Mach's principle. Based on  Mach's relational reality Barbour questioned the availability of constant external reference clocks and rods used to interpret GR and instead elevated the observation that rods can only be compared locally to a construction principle. This led Barbour (for a review see \cite{Barbour:2011dn}) to postulate relativity of spatial scale rather than relativity of simultaneity. In light of the success of spacetime ontology, this seems at first sight a foolish thing to do as it turns out that spatial scale invariance requires a fixed notion of simultaneity, which is incompatible with an integral part of relativity. Surprisingly however, one can abandon relativity of simultaneity and use the principle of local scale invariance alone to construct a theory of gravity that is {\it dynamically equivalent} to GR. We refer to this theory as Shape Dynamics (SD) \cite{Gomes:2010fh,Gomes:2011zi,Gomes:2011au}. 

Despite dynamical equivalence with GR, Shape Dynamics is at least as ``unnatural'' from a spacetime perspective as GR is from a Galilleian perspective. There are at least three points of contention: Shape Dynamics is (1) formulated as a purely Hamiltonian theory with (2) either a nonlocal Hamiltonian or frozen time and (3) refoliation invariance is incompatible with spatial scale invariance. It is the purpose of this paper to partially reconcile the spacetime ontology with the relational ontology of gravity. We identify the BRST formalism as a natural framework for such reconciliation  and then show that there is a hidden BRST symmetry in the GR action. This hidden BRST-invariance is due to Shape Dynamics, and arises from the fact that the GR action can be put on its head and be viewed as a gauge fixing for Shape Dynamics.  

\subsubsection*{Gauge Fixing and Shape Dynamics}

A heuristic explanation of why this scenario is at all possible goes as follows: gauge theories admit a description in terms of local fields by introducing redundant degrees of freedom. This redundancy is very useful, because it is in many cases the only way to provide a local description of a field theory. To get rid of this introduced  redundancy one demands that observables are invariant under the action of a group of gauge transformations $\mathcal{G}$. This means in the canonical formalism that one has a constraint surface $\Sigma$ that  is foliated into gauge orbits $\mathcal{O}_\alpha$, which turn out to be integrals of the Hamilton vector fields of first class constraints $\chi_\alpha$. Canonical observables are functions on the constraint surface that are constant along all gauge orbits and are thus uniquely defined by their values on a much smaller reduced phase space $\Sigma/\mathcal{O}_\alpha$ that contains only one point per gauge orbit. If this reduced phase space can be described as the intersection of the constraint surface (described by the constraints $\chi_\alpha$) with another first class constraint surface (described by the constraints $\sigma_\beta$), then each observable of the original system, described by $\chi_\alpha$, defines, under the assumption of some regularity conditions, a unique observable on a gauge theory that is described by the second first class surface $\sigma_\beta\approx 0$. This is particularly useful if the first class constraints $\sigma_\beta$ that define the second constraint surface are again local phase space functions. This explains how two incompatible symmetries can describe the same physics. 

The ADM constraints are due to the invariance of the Einstein-Hilbert action under spacetime diffeomorphisms and arise in the Legendre transform as secondary constraints, i.e. due to on-shell symmetries. This constraint algebra is abstractly the algebra of deformations of the surface of simultaneity under infinitesimal spacetime diffeomorphisms. The work of Hojman, Kucha\v{r} and Teitelboim \cite{HKT} shows that under strong locality assumptions the representation of the hypersurface deformation algebra on metric phase space is unique and coincides with the ADM constraints, which  thus  includes refoliation invariance, i.e. relativity of simultaneity. 

We understand, due to extensive work on the initial value problem of ADM by York and collaborators \cite{York:1973ia,York:1972sj}, that spatial conformal transformations can be used to gauge fix the refoliation constraints of ADM. Let us fix the notation already and call these $S(x)$ where $x$ is a continuous index representing spatial points.  The observation that spatial conformal transformations that preserve the total spatial volume are generated by a first class set of constraints - called here  ${D}(x)$ - allowed us to construct SD by trading all but one refoliation constraint of ADM for these conformal constraints. Schematically $S(x)\leftrightarrow  D(x) \cup H_{\mbox\tiny{gl}}$.

 By imposing constant mean curvature as a gauge fixing on ADM on the one hand and the refoliation constraints as gauge fixing for SD on the other hand, one sees that ADM and SD are dynamically equivalent, because it is possible to show that the two theories have identical initial value problem and identical equations of motion. In this paper we will , for the most part, go further and gauge fix ADM completely with a special  version of conformal constraints and a version of spatial harmonic gauge that is first class when combined with the conformal constraints. We refer to this non-dynamical theory as ``Extended Shape Dynamics'' (E-SD). {\it Symmetry trading} does not reconcile any two theories away form reduced phase space. This is where the BRST formalism comes in.

\subsubsection*{BRST and symmetry doubling}

The BRST formulation allows one to describe weak equalities, i.e. equalities that hold only on shell, through strong equations, i.e. equations that hold on the entire an extended phase space.  Moreover, the \emph{BRST-gauge-fixed Hamiltonian} is strongly invariant under BRST transformations, which is a rather counter-intuitive property for an object that has been gauge-fixed. The important observation for our purpose is the following: due to the existence of the second first class surface $\sigma_\beta$ that gauge fixes the original gauge symmetry $\chi_\alpha$, there is a second BRST transformation that leaves the \emph{gauge-fixed} Hamiltonian invariant. This in turn implies that the canonical Lagrangian is {\it invariant under these two BRST transformations}. We refer to this scenario as {\it symmetry doubling}.

We apply this symmetry doubling mechanism to GR in the ADM formulation.  The BRST-charge of ADM encodes foliation preserving spatial diffeomorphisms and refoliation invariance. 
We derive the E-SD BRST-charge using the standard construction but with the roles of ghosts and antighosts interchanged (i.e. we choose a preferred gauge-fixing fermion). We can then use the BRST-generator of E-SD as a gauge-fixing for ADM. It then follows from the vanishing of the on-shell Hamiltonian that this particular gauge-fixed ADM-action is invariant under both sets of BRST-transformations, justifying its name: ``Doubly General Relativity''.

Let us caution that the gauge-fixed BRST Hamiltonian (even at ghost number zero) is {\it not} the CMC-Hamiltonian with a particular version of spatial harmonic gauge, but turns out to be a distinctly different generator of true dynamics. We also note that the Hamiltonian is local, unlike the usual SD-Hamiltonian. Also, although we will provide a tentative interpretation of E-SD as a local version of the full conformal group, we think of it as Barbour's nondynamical scale- and diffeomorphism-invariant theory, because at the classical level one can trade the spatial conformal harmonic gauge constraints again for the equivalent set of diffeomorphism constraints and arrive precisely at this theory.

The appearance of a hidden invariance of the BRST-gauge fixed action may have many consequences. We will only discuss the most immediate ones in this paper: (1) Since the BRST-gauge-fixed action appears in the path integral, one can combine standard effective field theory reasoning with a semiclassical expansion to derive that the effective action should have the BRST-invariances of both ADM and E-SD in a semiclassical regime. This combined with dimensional analysis provides a construction principle for gravity theories that is different from general covariance. (2) The symmetry doubling mechanism appears to be generic for gravity theories and provides a bulk/bulk duality. The mechanism may thus provide insight into bulk/boundary dualities such as AdS/CFT. (3) Provided one has a solution to the quantum master equation, i.e. there are no quantum anomalies, one finds that the partition function for Doubly General Relativity is independent of the gauge fixing. So, viewing the partition function as a partition function for E-SD may open the opportunity to find gauge fixings other than ADM, which may improve power counting. (4) The algebra formed by the two distinct BRST charges is not anti-BRST \cite{AntiBRST}, but forms instead a supersymmetric algebra. 

\subsubsection*{Roadmap}

This paper is organized as follows: We start in section \ref{sec:canonical} with an analysis of the mechanism behind symmetry trading. We generalize the concept of Linking theory, as developed in \cite{Gomes:2011zi}, both for abelianized and non-abelianized constraints (as in Shape Dynamics.) We develop this in full generality. This section gives the state of the art of our work in canonical analysis. In the following section, section \ref{sec:E-SD} we apply this to gravity and show how the original SD theory  emerges. A reader familiar with the mechanism of symmetry trading and Shape Dynamics may want to skip this initial subsection. In the next subsection we construct E-SD. We briefly comment on the conformal constraint we require, which possesses a cosmological constant and has some special properties when acting on the momenta.  A rather involved analysis is required to find a gauge fixing for the spatial diffeomorphisms which has to remain first-class with respect to the conformal constraints.  In section \ref{sec:BRST} we start by performing the usual DeWitt-Fadeev-Popov construction for ADM with SD gauge-fixing. We comment on its short-comings for our own purposes, which is why we need to perform the full BRST construction. We then present a brief introduction to the  BRST-formalism and show how it applies to our case. Finally in section \ref{sec:DGR} we use all the previous results to construct Doubly General Relativity, where we also show the explicit form of the two BRST-transformations. In section \ref{sec:EFT} we apply standard effective field theory reasoning to obtain a revised definition of a gravity theory.  We briefly discuss some of the many possible consequences of Doubly General Relativity  before we conclude in \ref{sec:conclusions}.

\section{Setup: Equivalence of Gauge Theories}\label{sec:canonical}

Let us start this paper by reviewing and at the same time extending the results on the construction of equivalence of gauge theories obtained in \cite{Gomes:2011zi}. These results will form the basis for the construction in this paper. Let us first set the stage and fix notation.

\subsection{Canonical Setting}\label{sec:CanonSett}

We consider a canonical system $(\Gamma, \{.,.\},\{\chi_\alpha\}_{\alpha\in\mathcal A})$ on a symplectic phase space $\Gamma$ with canonical Poisson bracket $\{.,.\}$ and with a regular and irreducible set of first class constraints $\{\chi_\alpha\}_{\alpha\in\mathcal A}$, where the Hamiltonian $H$ is included in the first class constraints as an energy conservation constraint $H-E$. To make the argument more accessible, we will work with local expressions, rather than  global objects. For this we fix a local Darboux chart in $\Gamma$ and denote local coordinates by $(q_i,p^j)$ where $i,j$ range in $\mathcal I$ and where the only nonvanishing Poisson bracket of the coordinate functions is $\{q_i,p^j\}=\delta^j_i$. The constraint surface $\mathcal C=\{(q,p)\in \Gamma: \chi_\alpha(q,p)=0~\forall\alpha \in \mathcal A\}$ is by the first class assumption a coisotropic  surface in $\Gamma$. The Abelianization theorem says that one can find a local chart around around any point of the constraint surface where the first class constraints are given by $\chi_\alpha=q_\alpha$. 

It also follows that $\mathcal C$ is foliated by gauge\footnote{We assume that all first class constraints are generators of physical gauge transformations.} orbits whose infinitesimal generators are vector fields $v_\alpha: f \mapsto \{\chi_\alpha,f\}$. To select a unique representative from each gauge orbit, we impose canonical gauge fixing conditions $\{\sigma^\alpha\}_{\alpha \in \mathcal A}$, such that each gauge orbit intersects the gauge fixing surface $\mathcal G=\{(q,p)\in \Gamma:\sigma^\alpha(q,p)=0,~\forall\alpha\in\mathcal A\}$ exactly once and that the Dirac matrix $c_\alpha^\beta:=\{\chi_\alpha,\sigma^\beta\}$ is an invertible linear operator on $\mathcal P=\mathcal C\cap \mathcal G$. The initial value problem is then given by working out the reduced phase space $\mathcal P$ and the ``dynamics'' is encoded in the Dirac bracket on $\mathcal P$.

We will now impose that the gauge fixing conditions $\sigma^\alpha$ are a regular irreducible set of first class constraints themselves. This additional assumption will be very important subsequently but is not guaranteed: Take e.g. the first class constraints $\chi_1=q_1$, $\chi_2=q_2$ and the gauge fixing conditions $\sigma^1=p_1-f_1(q)$ and $\sigma^2=p_2-f_2(q)$, then $\{\sigma^1,\sigma^2\}=\partial_{q_1} f_2-\partial_{q_2} f_1$ which does not generically vanish when $\sigma^1=0=\sigma^2$. 

\subsubsection{Intermezzo: Intersecting Constraint Surfaces}

We will call two sets of constraint \emph{equivalent} if they define the same constraint surface on phase space. Qualities such as ``coisotropic" or  ``maximally symplectic"  hold for all equivalent constraints, since these are statements about the surfaces, and not their representatives. In this subsection, we need the following result: {\it Let $\gamma^\alpha$ be an irreducible set of regular first class constraints on the phase space $\Gamma$. If $\rho^\beta$ is a phase space functional that defines the same constraint surface $\gamma^\alpha=0$, then $\rho^\beta=M^\beta_\alpha\gamma^a$ for some invertible matrix valued phase space functional $M^\beta_\alpha$.}

Now, given a set of second class constraints $\{\sigma^\mu\}_{\mu \in \mathcal M}$, around each point in phase space one can find local Darboux coordinates $(q_i,p^j), ~i\in \mathcal{I}$ in  some open subset $U$ containing said point. There we have an invertible map $M^\nu_\mu$, such that $\{\tilde \sigma^\nu=M^\nu_\mu\sigma^\mu\}=\{q_\alpha,p^\alpha\}_{\alpha \in \mathcal A}$, so that the constraint surface $\{(q_i,p_i)\in \Gamma:\sigma^\mu(q,p)=0,~\forall \mu \in \mathcal M\}$ can be locally written as as the set $\{(q_i,p^j)\in U: q_\alpha=0=p^\alpha, ~\forall \alpha \in \mathcal A\}$, where $ \mathcal A\subset \mathcal{I}$. It is thus possible to write the second class surface locally as the intersection of the two first class surfaces defined by the constraint sets $\{q_\alpha=0\}_{ \alpha \in \mathcal A}$ and $\{p^\alpha=0\}_{\alpha \in \mathcal A}$. These two constraint sets are Abelian and canonically conjugates of each other, i.e. their Poisson bracket is $\{q_\alpha,p^\beta\}=\delta^\beta_\alpha$. We will now show that for any two regular, irreducible first class sets of constraints whose intersection is maximally second class it is always possible to find equivalent Abelian constraint sets that are canonical conjugates of one another.
\begin{prop}\label{prop:IntConstrSurf}
 Let $\{\chi^\alpha\}_{\alpha \in \mathcal A}$ and $\{\sigma_\alpha\}_{\alpha \in \mathcal A}$ be two sets of regular and irreducible first class constraints and let their intersection be maximally symplectic (i.e. $\{{\chi}^\alpha,{\sigma}_\beta\}$ is invertible in a neighborhood of the surface defined by ${\chi}^\alpha=0={\sigma}_\alpha$), and let $\{\eta^i\}_{i\in\mathcal{I}}$  be a set of first constraints also first class wrt both $\sigma$ and $\chi$. Then one can locally find two invertible maps $M^\alpha_\beta,N^\alpha_\beta$, such that the equivalent constraint sets $\{\tilde {\chi}^\alpha=M^\beta_\alpha {\chi}^\beta\}_{\alpha \in \mathcal A}$ and $\{\tilde {\sigma}^\alpha=N^\beta_\alpha {\sigma}^\beta\}_{\alpha \in \mathcal A}$ are Abelian and canonically conjugate (i.e. $\{{\tilde\chi}^\alpha,{\tilde\sigma}_\beta\}=\delta^\alpha_\beta$).
\end{prop}
{\bf proof:}\begin{enumerate}
             \item By the canonical representation theorem for first class constraints  (ex. 2.10 of \cite{HenneauxTeitelboim}) one can find a linear map $M^\alpha_\beta$ and a local Darboux chart on an open set $U$ with coordinates $(q_o^\alpha,\hat q^\mu,p^o_\beta,\hat p_\nu)$, whose only nonvanishing coordinate Poisson brackets are $\{q_o^\alpha,p^o_\beta\}=\delta^\alpha_\beta$ and $\{\hat q^\mu,\hat p_\mu\}=\delta^\mu_\nu$, such that the constraints $\tilde \sigma^\alpha=M^\alpha_\beta\sigma^\beta=q_o^\alpha$ are Abelian and equivalent to $\sigma^\alpha$. This amounts basically to finding coordinates on phase space adapted to the constraint surface $\sigma^\alpha=0$. 
             \item  $\frac{\partial {\chi}_\beta}{\partial p^o_\alpha}=\{{\chi}_\beta,\tilde \sigma^\alpha \}$ is still locally invertible on the constraint surfaces (maximally symplectic). Hence for each point on the constraint surface defined by the constraints ${\sigma}_\alpha$ there exists by the implicit function theorem an open neighborhood $\tilde U\subset U$ such that $\chi_\beta=0 $ if and only if $p^o_\beta=g_\beta(q_o,\hat q,\hat p) $. Thus  $\tilde {\chi}_\beta=p^o_\beta-g_\beta(q_o,\hat q,\hat p)$ is equivalent to ${\chi}_\beta$. 
             \item  The bracket $\{\tilde {\chi}_\alpha,\tilde {\chi}_\beta\}=(\partial_{q_o^\alpha}g_\beta-\partial_{q_o^\beta}g_\alpha)+\{g_\alpha,g_\beta\}$ by the first class property of the ${\chi}_\alpha$ vanishes at the constraint surface and  is independent of the $p^o_\beta$ everywhere.  Thus we have that  $\{\tilde {\chi}_\alpha,\tilde {\chi}_\beta\}$ vanishes throughout $\tilde U$. The same argument can be applied to the bracket $\{\eta^i, q_o^\alpha\}=-\frac{\partial {\eta}_i}{\partial p^o_\alpha}\approx 0$. 
             \item Since  ${\chi}_\alpha$ is regular and irreducible, by equivalence to $\tilde\chi^\alpha$ we know we  can find an invertible map $N_\beta^\alpha$, such that $\tilde {\chi}_\alpha=p^o_\alpha-g_\alpha=N^\alpha_\beta {\chi}_\beta$ (by the argument in ch. 5.2.1 in \cite{HenneauxTeitelboim}).
             \item $\{\tilde {\chi}^\alpha,\tilde {\sigma}_\beta\}=\{q_o^\alpha,p^o_\beta-g_\beta\}=\delta^\alpha_\beta, ~~\{\tilde {\chi}^\alpha,\eta^i\}\approx0$  and $\{\eta^i,\tilde {\sigma}_\beta\} \approx 0$. $\blacksquare$
            \end{enumerate}

A related difficulty is finding strong observables, which have to commute with the structure functions of the constraint algebra (by Jacobi identity). However, in a theory that has only structure constants one has locally a complete set of strong observables. We thus arrive at
\begin{prop}
 Let $\{{\chi}^\alpha\}_{\alpha \in \mathcal A}$ and $\{{\sigma}_\alpha\}_{\alpha \in \mathcal A}$ be two sets of regular and irreducible first class constraints and let their intersection be maximally symplectic (i.e. $\{{\chi}^\alpha,{\sigma}_\beta\}$ is invertible in a neighborhood of the surface defined by ${\chi}^\alpha=0={\sigma}_\alpha$), then there are two equivalent sets of first class constraints such that for each observable $B$ there exists an extension to a local phase space function that strongly commutes with all constraints.
\end{prop}
{\bf proof:} Observables are determined by a freely specifiable function $B$ on reduced phase space ${\chi}^\alpha=0={\sigma}_\beta$. Using the equivalent constraint sets of proposition \ref{prop:IntConstrSurf} we consider the vector fields $v_1^\alpha:f\mapsto \{{\tilde\chi}^\alpha,f\}$ and $v_2^\alpha:f\mapsto \{{\tilde\sigma}_\alpha,f\}$, so $[v_i^\alpha,v_j^\beta]:f\mapsto \{\epsilon_{ij}\delta^{\alpha\beta},f\}=0$, so by Frobenius' theorem one establishes local existence of a phase space function $A$ that commutes with the entire constraint set and satisfies the initial condition $A|_{{\chi}^\alpha=0={\sigma}_\beta}=B$. $\blacksquare$

\subsection{Construction of Equivalent Gauge Theories}

Assuming that the $\sigma^\alpha$ are first class allows us to use proposition \ref{prop:IntConstrSurf} to locally find two invertible maps $M^\alpha_\beta$ and $N^\alpha_\beta$, such that (1) the constraints $\tilde \chi_\alpha=M^\beta_\alpha \chi_\beta$ are Abelian and define the same constraint surface as the $\chi_\alpha$, (2) the $\tilde \sigma^\alpha=N^\alpha_\beta \sigma^\beta$ are Abelian and define the same constraint surface as the $\sigma^\alpha$ and (3) that $\{\tilde \chi_\alpha,\tilde \sigma^\beta\}=\delta^\beta_\alpha$. Moreover, we can locally find adapted Darboux coordinates $(q^o_\alpha,\hat q_\mu,p_o^\beta,\hat p^\nu)$, such that $\chi_\alpha=q^o_\alpha$ and $\sigma^\beta=p_o^\beta-g^\beta(q_o,\hat q,\hat p)$.  

Having an Abelian set of constraints $\{\tilde \chi_\alpha\}_{\alpha \in \mathcal A}$ and an Abelian set of gauge fixing conditions $\{\tilde\sigma^\alpha\}_{\alpha \in \mathcal A}$ with Poisson bracket $\{\tilde \chi_\alpha,\tilde \sigma^\beta\}=\delta^\beta_\alpha$ (as described in the previous subsection) on a phase space $(\Gamma,\{.,.\})$ allows one to construct a linking gauge theory. For this we extend phase space with a bosonic field $\phi_\alpha$ (for the BRST construction we will use a fermionic extension with ghosts) and canonically conjugate momentum $\pi^\alpha$. We then construct a set of  first class constraints on extended phase space from the set of second class constraints in the original phase space:
\begin{equation}\label{equ:AbelLinkConstr}
 \chi^1_\alpha=\phi_\alpha+\tilde \chi_\alpha,\,\,\,\,\chi^\alpha_2=-\pi^\alpha+\tilde \sigma^\alpha,\,\,\,\,\chi^i_3=\eta^i
\end{equation}
 Imposing the gauge condition $\phi_\alpha=0=\pi^\alpha$ an performing the phase space reduction yields the original system with second class constraints, proving equivalence of the extended system of constraints with the original one. In addition to this total gauge fixing there are two particularly interesting partial gauge fixings: $\phi_\alpha=0$ on the one hand and $\pi^\alpha=0$ on the other. 

We now {\bf define} the {\bf linking theory} (in the notation of \cite{Gomes:2011zi} a special linking theory) as the theory on the bosonically extended phase space as the system of Abelian first class constraints given by equation (\ref{equ:AbelLinkConstr}) together with the two sets of partial gauge fixing conditions $\phi_\alpha=0$ and $\pi^\alpha=0$. It follows
\begin{prop}
 The gauge theories $(\Gamma,\{.,.\},\{\tilde\chi_\alpha\}_{\alpha\in \mathcal A},\, \{\eta^i\}_{i\in\mathcal{I}})$ and $(\Gamma,\{.,.\},\{\tilde\sigma^\alpha\}_{\alpha\in \mathcal A},\, \{\eta^i\}_{i\in\mathcal{I}})$ are equivalent. 
\end{prop}
{\bf proof:} \begin{enumerate}
               \item Imposing the gauge fixing condition $\phi_\alpha=0$ implies that the constraints $\chi_2^\alpha$ are gauge-fixed since the Poisson-bracket $\{\phi_\alpha,\chi_2^\beta\}=\delta^\beta_\alpha$ is invertible. We thus perform the phase space reduction $(q_i,p^j;\phi_\alpha,\pi^\alpha)\to(q_i,p^j;0,\tilde \sigma^\alpha(q,p))$, which trivializes the constraints $\chi^\alpha_2=0$ and transforms the constraints $\chi^1_\alpha\to\tilde\chi_\alpha$. The Dirac bracket associated with this phase space reduction for functions $f,g$ on the non-extended phase space $\Gamma$ is $\{f,g\}_D=\{f,g\}+(\{f,\phi_\alpha\}\{\pi^\alpha-\sigma^\alpha,g\}-\{f,\pi^\alpha-\sigma^\alpha\}\{\phi_\alpha,g\})=\{f,g\}$, which coincides with the Poisson bracket on $\Gamma$. We thus obtain the gauge theory $(\Gamma,\{.,.\},\{\tilde\chi_\alpha\}_{\alpha\in \mathcal A},\, \{\eta^i\}_{i\in\mathcal{I}})$.
               \item Imposing the gauge fixing condition $\pi^\alpha=0$ and following the analogous steps as before yields the gauge theory $(\Gamma,\{.,.\},\{\tilde\sigma^\alpha\}_{\alpha \in \mathcal A},\, \{\eta^i\}_{i\in\mathcal{I}})$.
               \item Imposing the gauge fixing conditions $\tilde\sigma^\alpha=0$ to the gauge theory $(\Gamma,\{.,.\},\{\tilde\chi_\alpha\}_{\alpha\in \mathcal A},\, \{\eta^i\}_{i\in\mathcal{I}})$ and following the reduction procedure yields the same result as imposing the gauge fixing conditions $\tilde\chi_\alpha$ to the gauge theory $(\Gamma,\{.,.\},\{\tilde\sigma^\alpha\}_{\alpha \in \mathcal A},\, \{\eta^i\}_{i\in\mathcal{I}})$, since both cases coincide with imposing both conditions simultaneously $\phi_\alpha=0=\pi^\alpha$ and to the extended constraint set (\ref{equ:AbelLinkConstr}) and subsequently working out the reduced phase space and Dirac bracket.
             \end{enumerate}
We can thus find gauge fixing conditions for the two theories such that both have (1) identical initial value problems (i.e. identical reduced phase space) and (2) identical equations of motion (i.e. identical Dirac brackets on reduced phase space in the case where the Hamiltonian is replaced by an energy conservation constraint); we thus satisfy the definition of equivalent gauge theories given in \cite{Gomes:2011zi}. $\blacksquare$

\subsection{Nonabelian Constraints}

So far we have considered Abelianized constraints and canonically conjugate gauge Abelian fixing conditions. This is almost the most general case by proposition \ref{prop:IntConstrSurf}, which states that whenever one has a second class surface that is the intersection of two first class surfaces defined by ${\chi}_\alpha=0$ and ${\sigma}^\beta=0$ that gauge fix one another, one can locally find invertible linear maps $M^\alpha_\beta,N^\alpha_\beta$ such that the equivalent constraint sets $\tilde \chi_\alpha=M^\beta_\alpha\chi_\beta$ and $\tilde {\sigma}^\alpha=N^\alpha_\beta{\sigma}^\beta$ are each Abelian and satisfy $\{\tilde {\chi}_\alpha,\tilde {\sigma}^\beta\}=\delta^\beta_\alpha$.\footnote{Of course, given a general constraint system, an important part of the work is the actual separation of the constraints into its first class and maximally symplectic parts. } Let us now use this fact to apply the construction of equivalent gauge theories directly to the nonabelian case. For this we start with the bosonic phase space extension by adjoining canonically conjugate pairs $\phi_\alpha,\pi^\alpha$ to the original phase space and consider the constraints 
\begin{equation}\label{equ:NonAbelConstr}
 \hat {\chi}_\alpha=\phi_\alpha - \tilde\chi_\beta,\,~\textrm{ and }\,~\hat {\sigma}^\alpha= \pi^\beta + \tilde{\sigma}^\alpha.
\end{equation}
These constraints are equivalent to 
\begin{equation}
 \bar{\chi}_\alpha= (M^{-1})^\beta_\alpha\hat{\chi}_\beta=(M^{-1})^\beta_\alpha\phi_\beta-{\chi}_\alpha\,~\textrm{and}\,~\bar {\sigma}^\alpha = (N^{-1})_\beta^\alpha\hat{\sigma}^\beta=(N^{-1})^\alpha_\beta \pi^\beta+{\sigma}^\alpha,
\end{equation}
which are now expressed in terms of the nonabelian constraints $\chi_\alpha$ and $\sigma_\alpha$. It is trivial to check that the constraints $\hat {\chi}_\alpha$ and $\hat {\sigma}^\alpha$ all strongly commute on extended phase space $\Gamma_{ex}$. Using $\bar {\chi}_\alpha = (M^{-1})^\beta_\alpha\hat{\chi}_\beta$ and  $\bar {\sigma}^\alpha = (N^{-1})_\beta^\alpha\hat{\sigma}^\beta$ we find immediately that the constraints $\bar{\chi}_\alpha$ and $\bar {\sigma}^\alpha$ weakly commute, so the system is first class, since we assumed a generally covariant theory where the Hamitonian takes the form of an energy conservation constraint.

Then we impose the two sets of gauge fixing conditions $\phi_\alpha=0$ and $\pi^\beta=0$. We see immediately that the first gauge-fixes the constraints $\bar \sigma^\beta$, defining the extra bosonic variable as $\pi^\beta:=N^\alpha_\beta{\sigma}^\alpha$, and simplifies the constraints $\bar \chi_\alpha \to \chi_\alpha$ and the second gauge-fixes the constraints $\bar \chi_\alpha$ with $\chi^\beta:=M^\alpha_\beta{\chi}^\alpha$ and simplifies the constraints $\bar \sigma^\beta \to \sigma^\beta$.\footnote{As an illustration of how this happens in Shape Dynamics, where we are in fact dealing with non-abelianized constraints, we point out that the first gauge fixing defines the conformal momenta to $\pi_\phi:=4(\pi-\mean{\pi}\sqrt g)$, and the second fixes  the York conformal factor to $\phi:=\phi_o$. } The two phase space reductions are thus 
\begin{equation}
  \begin{array}{rcl}
    (\phi_\alpha,\pi^\beta) &\to&(0,\sigma^\beta)\\ 
    (\phi_\alpha,\pi^\beta) &\to&(\chi_\alpha,0).
  \end{array}
\end{equation}
The Dirac-bracket associated with these two  phase space reductions of $\Gamma_{ex}$ coincides with the Poisson-bracket on $\Gamma$.  We can thus define the special {\bf linking theory} in the nonabelian case as the gauge theory $(\Gamma_{ex},\{.,.\},\{\bar \chi_\alpha,\bar \sigma^\alpha\}_{\alpha \in \mathcal A},\, \{\eta^i\}_{i\in\mathcal{I}})$ together with the two partial gauge fixing conditions $\{\phi_\alpha\}_{\alpha\in \mathcal A}$ and $\{\pi^\alpha\}_{\alpha\in \mathcal A}$. 

To retain the generality that we had in the previous section, we would like to include constraints $\{\eta^i\}_{i\in\mathcal{I}}$ that are not traded. It turns out that, with some extra notational effort, one can follow the same construction as the one described in this subsection when one adjoins the additional constraints to the original constraints on both sides of the construction. The proof of the following theorem is then analogous to the Abelian case:
\begin{prop}\label{prop:general_sym_trading}
  The gauge theories $(\Gamma,\{.,.\},\{\chi_\alpha\}_{\alpha\in \mathcal A},\, \{\eta^i\}_{i\in\mathcal{I}})$ and $(\Gamma,\{.,.\},\{\sigma^\alpha\}_{\alpha\in \mathcal A},\, \{\eta^i\}_{i\in\mathcal{I}})$ are equivalent also in the nonabelian case. 
\end{prop}

\subsection{Relation of Observable Algebras}

Unlike the BRST formalism where one works with strong equations for observables, one works with weak equations in the linking theory formalism. One thus needs to work out the relation between the observables of the two equivalent gauge theories. The most straightforward way to obtain such a relation is to work out the observable algebra in the linking theory: Let us assume that we have constructed the Poisson-algebra of weak observables, i.e. we have a complete set of phase space functions $A$ on $\Gamma_{ex}$ that weakly commute with all constraints of the linking theory, so the Jacobi-identity implies that $\{A_1,A_2\}$ is also a weak observable. 

We can now use that the phase space reduction is a morphism of the Poisson algebra, since the Dirac bracket coincides with the Poisson bracket, to identify the observables obtained by the two phase space reductions with one another. This identification $I$ reads explicitly
\begin{equation}
  I: A|_{\phi=\phi_o,\pi=0} \leftrightarrow A|_{\phi=0,\pi=\pi_o}.
\end{equation}
This identification is of course not one to one in terms of phase space functions, but only one to one between the equivalence classes of phase space functions that define the same observable. $I$ is then a morphism of the observables algebras, when viewed as a map between these equivalence classes .

A significant simplification occurs in the case of two sets Abelian constraints $\tilde \chi^1_\alpha$ and $\tilde \chi_2^\beta$ with Poisson bracket  $\{\tilde \chi^1_\alpha,\tilde \chi_2^\beta\}=\delta^\beta_\alpha$. In this case there exists, by Frobenius' theorem, locally in phase space $\Gamma$ a complete set of phase space functions $\tilde A$ that strongly commutes with both sets of constraints. These are special representatives $\tilde A$ of the equivalence classes that define observables, now viewed as phase space functions on $\Gamma_{ex}$ that are independent of $\phi_\alpha,\pi^\beta$, which are strong observables for the linking theory. The identification among these observbales is one to one and is given by the identity map
\begin{equation}
  I : \tilde A \leftrightarrow \tilde A.
\end{equation}

\section{Shape Dynamics and Extended Shape Dynamics}\label{sec:E-SD}

In this section we will give a brief introduction to Shape Dynamics as it is usually presented, and then present the extended version, which completely gauge-fixes ADM while still forming a first class system of constraints. The reader familiar with the construction of Shape Dynamics may want to skip the first part. 

\subsection{Shape Dynamics}
Let us now briefly review the recipe for the construction of Shape Dynamics as a theory equivalent to ADM gravity on a compact Cauchy surface $\Sigma$ without boundary. For details see \cite{Gomes:2011zi}.
\begin{enumerate}
\item Start with the standard ADM phase space $\Gamma_{ADM}=\{(g,\pi):g\in \mathrm{Riem},\pi\in T_g^*(\mathrm{Riem})\}$, where $\mathrm{Riem}$ denotes the set of Riemannian metrics on the above defined 3-manifold $\Sigma$, and the usual first class ADM constraints, i.e. the scalar constraints $S(x)=\frac{\pi^{ab}\pi_{ab}-\frac{1}{2}\pi^2}{\sqrt g}-\sqrt g R$ and momentum constraints $H^a(x)=\pi^{ab}_{~;b}(x)$ thereon. 
\item Extend the ADM phase space with the phase space of a scalar field $\phi(x)$ and its canonically conjugate momentum density $\pi_\phi(x)$, which we introduce as additional first class constraints $\mathcal{Q}(x)=\pi_\phi(x)\approx 0$. The system is thus merely a trivial embedding of the original ADM onto the extended phase space. 
\item Now  use the canonical transformation $T_\phi$ generated by the generating functional $F=\int d^3x\left(g_{ab}e^{4\hat \phi}\Pi^{ab}+\phi\Pi_\phi\right)$,  where $\hat \phi(x):=\phi(x)-\frac 1 6 \ln\langle e^{6\phi}\rangle_g$ using the mean $\langle f\rangle_g:=\frac 1 V \int d^3x\sqrt{|g|} f(x)$ and 3-volume $V_g:=\int d^3x\sqrt{|g|}$. The triangle brackets - which we call ``means" -  and the volumes technically appear as a manner to yield dynamically equivalent theories. They force the appearance of a global Hamiltonian in SD. The effect of the generating functional on the canonical variables is:  
\be\begin{array}{rcl}
    t_\phi g_{ab}= e^{4 \hat \phi(x)} g_{ab}&,& t_\phi\pi^{ab}=e^{-4 \hat \phi} \left(\pi^{ab}-\frac 1 3 \langle \pi\rangle_g \left(1-e^{6\hat \phi}\right)g^{ab}\sqrt{|g|}\right)\\
    t_\phi \pi_\phi=\pi_\phi-4\left(\pi(x)-\sqrt{g}(x)\langle\pi\rangle_g\right)&,& t_\phi \phi'= \phi+\phi'. 
\end{array}\ee
 At this point you should have the first class set of constraints
$    S(t_\phi g_{ab}(x), t_\phi \pi^{ab}(x)),~
    H(t_\phi g_{ab}(x), t_\phi \pi^{ab}(x))$ and $
   t_\phi Q(x)=\pi_\phi(x)-4(\pi-\langle\pi\rangle\sqrt g)(x)$. 
\item Perform the phase space reduction for the gauge fixing condition $\pi_\phi(x)=0$.
\end{enumerate}

After some algebra, we find  that all but one of the $  t_\phi S(x)$ constraints can be solved for $\phi=\phi_o$, and this one leftover constraint yields the global Hamiltonian of SD. The technical requirement of this assertion is basically the same as that required by the existence and uniqueness theorems for the Lichnerowicz-York equation with the extra demand that the conformal factor be volume-preserving, which is what requires one global Hamiltonian to remain unsolved.\footnote{ We should also mention the fact that the inclusion of the volume-preserving condition extends the solvability of the Lichnerowicz-York equation beyond the case of Yamabe negative metrics, because it is a canonical way to impose York-scaling.}  This global Hamiltonian can be seen to reflect exactly a constraint on the total volume of the Universe.

 The end result is the following set of first class constraints, which define Shape Dynamics
\begin{equation}\label{equ:pureSDconstraints}
 \begin{array}{rcl}
   H_{\mbox{\tiny{SD}}}&=&V-\int d^3 x\sqrt g e^{6\phi}\\
   H^a(\xi_a)&=&\int d^3x \pi^{ab}\mathcal{L}_\xi g_{ab}\\
   D(\rho)&=&\int d^3x \rho\left(\pi-\langle \pi\rangle\sqrt{|g|}\right),
 \end{array}
\end{equation}
where $\phi_o(g,\pi)$ is such that $H_{\mbox{\tiny{SD}}}=0$ combined with $\phi=\phi_o(g,\pi)$ is equivalent to $T_\phi S(x)=0$ (at the surface $\pi_\phi=0$). 

We should also mention that the only other gauge-fixing of the conformal variables that works as a gauge-fixing (has the required solvability properties) is $\phi=0$, and this yields back ADM. The main accomplishments of Shape Dynamics is that it is a true gauge theory for the conformal factor, and it doesn't require non-localities either in the canonical variables (as for example the usual transverse traceless momenta of the York method demand), or in the local constraints, which is what one gets when one tries to use the constant mean curvature gauge in usual ADM, but without solving the momentum constraint \cite{Isham}.

In the language of section \ref{sec:canonical}, the construction present here would amount to a different route to the same result. Let us use this opportunity to show the derivation of a slightly different result, the construction of maximal Shape Dynamics, with $\pi-\lambda\sqrt g$ as the gauge-fixing, instead of $\pi-\mean{\pi}\sqrt g$. This is the model used in section \ref{sec:E-SD}.

We will follow the construction necessary for proposition \ref{prop:IntConstrSurf}. We start with the constraints $S(x)=0$ playing the role of the $\chi^\alpha$ constraints, and $\pi-\lambda\sqrt g=0$ playing the role of $\sigma^\beta$.

One then decomposes the standard gravitational canonical variables as 
$$(g_{ab},\pi^{ab})\mapsto (-\sqrt g,\frac{ 2\pi}{3\sqrt g}, g^{-1/3} g_{ab}, g^{1/3}\sigma^{ab})=(P,T, \gamma_{ab},\rho^{ab})$$
where $\sigma^{ab}$ is the traceless part of $\pi^{ab}$. It is possible to show that $P,T$ and $\gamma_{ab}, \rho^{ab}$ form  canonical pairs, and the first pair commutes with the second pair.   The constraint $T-\lambda=0$ is abelian. So we are already at a point where we have identified $T$ with the  $q^\alpha_o$ in the first step of proposition \ref{prop:IntConstrSurf}. The Poisson bracket between $T$ and $S(x)$ is invertible, as we show later in section \ref{sec:E-SD}, in lemma \ref{lem:gf_scalar}. One now sees that it is possible to rewrite the scalar constraint $S(x)=0$  as $P=F[\gamma_{ab}, \rho^{ab}, T]$. The rest of the propositions and results of section \ref{sec:canonical} follow.

\subsection{Extended Shape Dynamics}

The construction of standard Shape Dynamics is complicated by the fact that we demand that it generate dynamical trajectories equivalent to ADM. In the present case, we will drop this requirement and merely construct a covariant dynamical system which defines a coisotropic surface  (i.e. one that is composed of first class constraints) and that is maximally symplectic with respect to the ADM constraints. In the case of total symmetry trading and a completely covariant system, as we have mentioned, this condition is enough to guarantee observable duality. Unlike the construction of Shape Dynamics, we do not need to introduce a Linking Theory, because all we will care about for the BRST treatment is that, formally $\det{|\{\chi_{\mbox{\tiny{ ADM}}}, \sigma_{\mbox{\tiny{ E-SD}}}\}|}\neq 0$. Thus we avoid global properties of the gauge-fixing surfaces.

\subsubsection{The conformal gauge}

We start by presenting the conformal constraint and some of its properties. We define the modified spatial conformal transformation generator as $Q(x):=\pi(x)-\lambda\sqrt g$. It is modified in the sense that the transformation on the metric acts as  $\{\pi(\rho),g_{ab}(x)\} =\rho g_{ab}$ and on the metric momenta acts as  $\{Q(\rho),\pi^{ab}(x)\} =-\rho(\pi^{ab}-\frac{\lambda}2 g^{ab}\sqrt g)$. This transformation can not be included as some restriction on the conformal factor, since the metric transforms fully under Weyl transformations. But let us take a closer look on its action on the traceless part of $\pi^{ab}$, i.e. on $\sigma^{ab}:=\pi^{ab}-\frac{1}{3}\pi g^{ab}$:  
$$\{Q(\rho),\pi^{ab}-\frac{1}{3}\pi g^{ab}\} =-\rho(\pi^{ab}-\frac{1}{3}\pi g^{ab})$$
and hence it acts as conformal transformation on the traceless part of $\pi^{ab}$. The generating functional for this transformation would be $F=\int d^3x\left(g_{ab}e^{4\phi}\Pi^{ab}+\lambda V\right)$. This is what one would in fact expect from a conformal transformation that includes the cosmological constant: it deforms only the pure trace part of $\pi^{ab}$, and only by a pure trace deformation. 

 Clearly $Q(x)=0, \forall x\in M$  forms a set of first class constraints, and furthermore the conformal gauge $Q(x)$ forms a weakly invertible Poisson bracket with $S(x)=0$, which we now verify:
\begin{equation}\{S(N),Q(x)\}=\left(-\frac{3}{2}S(x)+2\sqrt g( \nabla^2 -R-\frac{1}{4\sqrt g}\lambda\pi))\right)N(x)
\end{equation}
Thus on the constraint surface $Q(x)\approx 0, S(x)\approx 0$ this becomes 
\begin{equation}\{S(N),\pi(x)\}\approx 2\sqrt g( \nabla^2 -\bar\sigma_{ab}\bar\sigma^{ab}(x)-\frac{1}{12}\lambda^2)N(x)
\end{equation}
where $\bar\sigma^{ab}$ is the undensitized traceless part of the momenta. The operator $\nabla^2 -\bar\sigma_{ab}\bar\sigma^{ab}(x)-\frac{1}{12}\lambda^2$ has a strictly negative linear part, and thus by the maximum principle is invertible. Thus the Poisson bracket is invertible.
We can thus state the:
\begin{lem}\label{lem:gf_scalar}
The scalar constraint $S(x)=0$ and the deformed conformal constraint $Q(x)=0$ form a weakly invertible Poisson bracket. 
\end{lem}

 Now we move on to the second constraint that will help compose the ``Shape Dynamics" BRST gauge-fixing fermion.

\subsubsection{Conformal Harmonic Gauge through harmonic mappings}

Let us first produce full symmetry trading, i.e. we will first construct a gauge-fixing of both the scalar- and diffeomorphism- constraints that is itself first class.  This is what we now turn to. Later we will return to a partial gauge fixing trading only the scalar constraints of ADM for conformal constraints of SD.

To preform symmetry trading, we need two first class systems whose intersection is second class. In particular, we want to gauge-fix the ADM constraints
\begin{equation}
 \begin{array}{rcl}
  S(N)&=&\int d^3x N \left(\frac{1}{\sqrt{g}} \pi^{ab}G_{abcd}\pi^{cd}-R\right)\\
  H(u)&=&\int d^3x \pi^{ab} \mathcal L_u g_{ab}
 \end{array}
\end{equation}
with the parametrized constant mean curvature + spatial conformal harmonic gauge (PCMC-SCHG) conditions
\begin{equation}
 \begin{array}{rcl}
  Q(x)&=&g_{ab}(x)\pi^{ab}(x)-\lambda \sqrt{g}(x)\\
  F^k(x)&=&(g^{ab}\delta^k_c+\frac 1 3 g^{ak}\delta^b_c) e^c_\alpha(\nabla_a-\hat \nabla_a)e^\alpha_b,
 \end{array}
\end{equation}
which are defined using the background metric $\hat g$. It is inevitable that we use \emph{some}  fixed background structure as it is exactly the 3-diffeomorphisms that we are trying to gauge-fix.

 Using that $\hat g$ (all hatted symbols are derived from $\hat g$) is a phase space constant we see immediately $\{Q(x),Q(y)\}=0=\{F^i(x),F^j(y)\}$. Using the local expression
\begin{equation}\label{equ:F_a}
 F^k=(g^{ab}\delta^k_c+\frac 1 3 g^{ak}\delta^b_c)\Delta^c_{ab},
\end{equation}
where we the tensor $\Delta^c_{ab}=\Gamma^c_{ab}-\hat \Gamma^c_{ab}$ is obtained as the difference of two connections and is symmetric in its lower indices. Using that under local conformal transformations the Christoffel symbols transform as:
$$T_\phi \Gamma^k_{ij}=\Gamma^k_{ij}+\frac 1 2 (\delta^k_i\phi_{,j}+\delta^k_j \phi_{,i}-g_{ij}g^{kl}\phi_{,l})$$
we find that
\begin{equation}
 \{F^k(x),Q(y)\}=\left.\frac{\delta}{\delta \phi(y)} F^k(x)\right|_{\phi\equiv 0}=-\delta(x,y) F^k(x),
\end{equation}
which shows
\begin{lem}
 The system formed by the intersection of the {\bf parametrized constant mean curvature} (PCMC) constraint and the {\bf spatial conformal harmonic gauge fixing} (SCHG) is first class.
\end{lem}

To have a proper local (in phase space) gauge fixing, or equivalently a second class intersection of the two first class surfaces, we need to show that the operator 
\begin{equation}
 K:
\left(\begin{array}{c}
u(x)\\
N(x)\end{array} \right) \mapsto
 \left(\begin{array}{c}
\{F^k(x),H(u)+S(N)\}\\
\{Q(x),H(u)+S(N)\}\end{array} \right)
\end{equation}
is weakly (i.e. on the surface $Q=F=H=S=0$) an isomorphism. We can write $K$ schematically as:
\begin{equation}
\left(\begin{array}{rl}
\{Q, S\}& \{Q,H\}\\
\{F, S\}&\{F,H\}
\end{array}\right)
\end{equation}

Since, $\{Q(x),H(u)\}=\mathcal L_u Q(x)$,  $K$ operator matrix is weakly triangular, so it suffices to show the operators
\begin{equation}
 \begin{array}{rcl}
  (Au)^k: x &\mapsto& \{F^k(x),H(u)\}\\
  BN: x &\mapsto& \{Q(x),S(N)\}
 \end{array}
\end{equation}
are weakly isomorphisms. In the following we will follow the strategy of \cite{Andersson:2001kw} and use 
$$\{F^k,H(u)\}=\{g^{ab}\delta^k_c+\frac 1 3 g^{ak}\delta^b_c,H(u)\}\Delta^c_{ab}+(g^{ab}\delta^k_c+\frac 1 3 g^{ak}\delta^b_c)\{\Gamma^c_{ab},H(u)\}=\mathcal L_u F^k + (g^{ab}\delta^k_c+\frac 1 3 g^{ak}\delta^b_c)\mathcal L_u \hat \Gamma^c_{ab}$$
 and
\begin{equation}
 \mathcal L_u \hat \Gamma^c_{ab}=\frac 1 2 \left[(\hat {R^c}_{adb}+\hat {R^c}_{bda})u^d+(\hat \nabla_a \hat \nabla_b+\hat \nabla_b \hat \nabla_a)u^c\right]=\hat \nabla_a \hat\nabla_b u^c+\hat {R^c}_{bda}u^d
\end{equation}
we find that $A$ takes weakly (i.e. after subtracting $\mathcal L_u F^k$) the local form
\begin{equation}
 (Au)^k=(g^{ab}\delta^k_c+\frac 1 3 g^{ak}\delta^b_c)(\hat \nabla_a\hat \nabla_b u^c +\hat {R^c}_{bda}u^d).
\end{equation}
The leading symbol of $A$ is $||\xi||^2_g\delta^k_a+\frac 1 3 \xi^k \xi_a$, which has the inverse $||\xi||_g^{-2}\delta^a_l-\frac 1 4 ||\xi||_g^{-4}\xi^a\xi_l$ for every nonvanishing $\xi$. The operator $A$ is thus elliptic and we can us The Fredholm alternative for elliptic operators, which limits both the kernel of $A$ and of its adjoint $A^*$ to be finite-dimensional. In other words, we have 
$$A:H^{s+1}=\mbox{Ker}(A)\oplus \mbox{Im}(A^*)\rightarrow \mbox{Im}(A)\oplus \mbox{Ker}(A^*)=H^{s-1}
$$
Thus we can proceed as if we were working with a finite-dimensional linear algebra to deduce from $\langle u,u\rangle >0 \,\Rightarrow \langle u,A u\rangle<0$ that $A$ has trivial kernel and cokernel, and is thus an isomorphism between the Sobolev spaces $H^{s+1}$ and $H^{s-1}$. But even if we require no such condition on $A$ having trivial kernel and co-kernel, for our purposes it would still be suitable to work with a finite-dimensional one. This is because this condition already precludes the Hessian of the action from being degenerate. Moreover, such poles could possibly have physical meaning, as is usually the case. Nonetheless, we will find conditions on our auxiliary background metric that realizes a trivial kernel and cokernel for $A$. 

 This is not an easy task,\footnote{The authors spent a couple of months attempting different strategies.} and as far as we know the only way forward is to use the notion of \emph{harmonic mappings}, adopted from  \cite{Andersson:2001kw} (see \cite{Wood} for a more thorough introduction to the use of harmonic mappings). We will start by introducing  the connection $D$ along a map $\phi:(M,g)\rightarrow(N,\hat g)$ (using Latin indices on $M$ and Greek indices on $N$) with local action
\begin{equation}
 \begin{array}{rcl}
   D_i t^\gamma&=&t^\gamma_{,i}+\hat \Gamma^\gamma_{\alpha\beta}t^\alpha \phi^\beta_{,i}\\
   D_i t^\gamma_j&=&t^\gamma_{j,i}-\Gamma^k_{ij}t^\gamma_k+\hat \Gamma^\gamma_{\alpha\beta}t^\alpha_j \phi^\beta_{,i}.
 \end{array}
\end{equation}
This connection is metric compatible w.r.t. the inner product on $\otimes^k T^*(M)\otimes \phi^*T(N)$ given by
$$\int d\mu(g)\langle t,\tilde t\rangle =\int d\mu(g)g^{i_1j_1}...g^{i_kj_k}h_{\alpha\beta}t^\alpha_{i_1...i_k}\tilde t^\beta_{j_1...j_k}$$
It allows us to have metric compatibility with two different metrics: one for covariant indices and one for contravariant ones. This is the property that makes the use of Harmonic mappings so powerful in the context of using a background metric.   The Laplacian $\Delta_D t^\gamma=g^{ab}D_aD_bt^\gamma$ is self-adjoint w.r.t. this inner product since 
$$\int d\mu(g)\langle \Delta_D u,v\rangle=-\int d\mu(g)\langle D u,Dv\rangle$$

 Let us then in particular consider $M=N$ and $\phi=Id$, so $\phi^\alpha_i=\delta^\alpha_i$, so we can write
\begin{equation}
 \begin{array}{rcl}
   (g^{ij}\delta^k_c+\frac{1}{3} g^{ik}\delta^j_c)\hat \nabla_i \hat \nabla_j u^c&=&(g^{ij}\delta^k_c+\frac{1}{3} g^{ik}\delta^j_c)(\partial_i \hat \nabla_j u^c-\hat \Gamma^r_{ij}\hat \nabla_r u^c+\hat \Gamma^c_{ir}\hat \nabla_r u^r)\\
      &=&(g^{ij}\delta^k_c+\frac{1}{3} g^{ik}\delta^j_c)(\partial_i \hat \nabla_j u^c-\Gamma^r_{ij}\hat \nabla_r u^c+\hat \Gamma^c_{ir}\hat \nabla_r u^r)+W^k\\
      &=&(g^{ij}\delta^k_c+\frac{1}{3} g^{ik}\delta^j_c)D_iD_ju^c+W^k,
 \end{array}
\end{equation}
where $W^k=(g^{ij}\delta^k_c+\frac{1}{3} g^{ik}\delta^j_c)\Delta^r_{ij}\hat \nabla_r u^c$, which does not weakly vanish unlike the usual non-conformal harmonic gauge. The condition $\langle u,Au\rangle<0$ thus translates into
\begin{equation}
 0 > \int d\mu(g) \hat g_{lk} u^l (\Delta_D u^k+\frac{1}{3} g^{kj}D_jD_iu^i + g^{ab}\hat {R^k}_{bda}u^d + (g^{ij}\delta^k_c+\frac{1}{3} g^{ik}\delta^j_c)\Delta^r_{ij}\hat \nabla_r u^c).
\end{equation}
Estimating $\int d\mu(g) \langle u, \Delta_D u\rangle=-\int d\mu(g) \langle Du, Du\rangle<0$. We still want to show that   
$$\int d\mu(g)\frac{1}{3} u^l \hat g_{lk}  g^{kj}D_jD_iu^i<0$$
 or stipulate conditions under which it is always negative, which would automatically be true if $ \hat g_{lk}  g^{kj}=\delta^j_l$. An equivalent positivity condition is to assume that the  matrix
\be\label{equ:diagon} M^j_l= \hat g_{lk}  g^{kj}-\delta^j_l
\ee is diagonalizable and has  non-negative eigenvalues. 

Which then  allows us to then restrict our attention to 
\begin{equation}
 \int d^3x \sqrt{|g|}\hat g_{kl}u^l (g^{ij}\delta^k_c+\frac 1 3 g^{ik}\delta^j_c)\Delta^r_{ij}(\hat \nabla_r u^c)=-\frac 1 2 \int d^3x\sqrt{|\hat g|} U_{lc} u^lu^c,
\end{equation}
where 
$$U_{lc}=\hat \nabla_r\left(\sqrt{\frac{|g|}{|\hat g|}}\hat g_{kl}(g^{ij}\delta^k_c+\frac 13 g^{ik}\delta^j_c)\Delta^r_{ij}\right)$$
 and where we integrated by parts and used the symmetry under the interchange $l\leftrightarrow c$. Thus finally we are able to get a condition that does not require any special behavior from the smearing functions, and is just a condition on the auxiliary metric. 
This turns the condition into 
\begin{equation}
  g^{ab}\hat g_{lk}{\hat R^k}_{acb} < \frac 1 2 U_{lc},
\end{equation}
where the inequality means that at each point the linear map $g^{ab}\hat g_{lk}{\hat R^k}_{acb} - \frac 1 2 U_{lc}$ is strictly negative (i.e. in a local chart the matrix has three negative eigenvalues).
 It is not hard to show that these conditions have non-empty intersection.

 For example, if $g_{ij}$ has negative Ricci curvature (i.e. eigenvalues of $R_{ij}$ are negative) then one can take $\hat g_{ij}$. And since the conditions on $\hat g_{ij} $ are open (i.e. inequalities) there exists always an open set once you have satisfied the conditions. But of course, $\hat g_{ij}$ is a fixed background, and thus this suffices to show only that for each $g_{ij}$ there exists a $\hat g_{ij}$ that makes the problem locally well-posed. As further investigation of these conditions would take us even farther afield, we avoid it here.

 We have thus established
\begin{lem}\label{lem:AIsomorphism}
 If $g^{ab}\hat g_{lk}{\hat R^k}_{acb} < \frac 1 2 U_{lc}$ as a linear map at every point in a Riemannian manifold without boundary, and $M^i_{\phantom{j}j}$ given in \eqref{equ:diagon} is diagonalizable and with positive eigenvalues, then $A$ is an isomorphism between $H^{s+1}$ and $H^{s-1}$.
\end{lem}

\begin{lem}
  If the same conditions hold as in lemma \ref{lem:AIsomorphism} then the operator $K$ is an isomorphism between $H^{s+1}$ and $H^{s-1}$.
\end{lem}
Using Lemma \ref{lem:gf_scalar} now have two first class surfaces (ADM and PCMC-SCHG) whose intersection is locally second class, yielding
\begin{theo}
 The ADM system and the PCMC-SCHG system are equivalent in the part of phase space where $g^{ab}\hat g_{lk}{\hat R^k}_{acb} < \frac 1 2 U_{lc}$ (as a linear map everywhere on a compact Riemannian manifold without boundary) and $M^i_{\phantom{j}j}$ given in \eqref{equ:diagon} is diagonalizable and with positive eigenvalues.
\end{theo}
It is of course disappointing that the equivalence does not generically hold, and so many extra conditions are required for the argument to hold, allowing us to say . 

However, in the course of establishing Lemma \ref{lem:AIsomorphism} we also established that $A$ has a finite dimensional kernel. Moreover, symmetry doubling in the effective field theory sense is on the other hand still possible, because the BRST-Hamiltonian will still have two invariances. A finite dimesnional kernel however means that the observable algebras may mismatch by precisely the the elements of the kernel.

\subsubsection*{Conformal symmetry and Weyl symmetry}

What we have been here referring to as conformal symmetry is usually referred to as Weyl symmetry. 
The QFT conformal symmetry group in a flat background plays an important role in the construction of quantum filed theories. Its Lie algebra is obtained by adjoining the dilatation generator $D$ and the special conformal generators $K_a$ to the (Euclidean/Lorentzian) Poincar\'e generators $M_{ab}$, $P_a$. The fundamental representation of this group as vector fields on Euclidean/Minkowski space is
\begin{equation}\label{equ:ConformalGenerators}
 \begin{array}{rclcrcl}
  P_a&=&-\partial_a,&&M_{ab}&=&x_a\partial_b-x_b\partial_a,\\
  D&=&-x^a\partial_a,&&K_a&=&x^2\partial_a-2x_ax^b\partial_b.
 \end{array}
\end{equation}
If we go to a chart $(U,\phi)$ then we can choose these vector-fields as smearing functions for the spatial diffeomorphism constraint and obtain a local representation of this Lie-algebra. This is however {\bf not} what we are seeking, because we want is a {\bf local} representation of the sub-algebra generated by $D$ and $K_a$ where in our case $D(x)$ acts on metric functions as $g_{ab}\diby {}{g_{ab}}$, because this $D(x)$ generates local scale transformations in a general background. Of course, on phase space we want to replace this $D(x)\rightarrow Q(x)$, the PCMC. We thus seek local generators $K_a(x)$ that satisfy the Poisson-brackets $\{Q(x),K_a(y)\}=-K_a(x)\delta(x,y)$ and $\{K_a(x),K_b(y)\}=0$. Thus we require the smeared Poisson brackets
\begin{equation}
 \begin{array}{rclcrcl}
   \{K(u),K(v)\}&=&0,&&\{Q(\rho),K(v)\}&=&-K(\rho v),
 \end{array}
\end{equation}
where $\rho$ is a scalar function and $u,v$ are vector fields and where $(\rho v)^a(x)=\rho(x)v^a(x)$. There is an infinite number of distinct $K_a(x)$ that satisfy these relations; here we made a very simple choice that is in particular independent of the metric momentum, which implements the relation $\{K_a(x),K_b(y)\}=0$ from the onset: $K_a=F_a$, given in equation \eqref{equ:F_a}.

We also note that if our manifold is conformally flat, i.e. has a maximal set of conformal Killing vector fields, we automatically incorporate in our framework the equivalent of special conformal transformations. They are  left as a subset of the spatial diffeomorphisms that the $F^a$ do not fix and are thus remaining symmetries of our theory. This framework also allows us to have non-maximal sets of conformal Killing vector fields.

\section{BRST Construction}\label{sec:BRST}

\subsection{De Witt Fadeev-Popov}

The local constraints of Shape Dynamics have a very simple geometric meaning: they are generators of spatial diffeomorphisms and spatial conformal transformations that do not change the total spatial volume. This enables us to find the explicit gauge orbits. The price for this simple form of the local constraints is a complicated time-reparametrization constraints which turns out to be nonlocal. This nonlocality disappears when the time-reparametrization of Shape Dynamics constraint is restricted to the surface where all scalar constraints of ADM gravity except for one hold. The restriction of the time-reparametrization constraint of Shape Dynamics to this surface is proportional to the one scalar constraint that is not imposed. This is a very benign nonlocality from the point of view of a canonical path integral, because to define this path integral one needs to impose gauge fixing conditions, which may be chosen to be the ADM constraints. Let us now consider the formal\footnote{We use the term \lq\lq{}formal\rq\rq{} to point out that we have no reason to believe that this path integral is well defined.} path integral for the partition function of Shape Dynamics
\begin{equation}
Z=\int Dg_{ab}\,D\pi^{ab} \delta[\chi] \delta[\sigma] |det(\{\chi,\sigma\})| \exp\left(i\int dt\,d^3x\, \pi^{ab}\dot g_{ab}\right),
\end{equation}
where $\delta[\chi]:=\prod_{\alpha}\delta[\chi^\alpha]$ denotes the product over delta functionals of all constraints and $\delta[\sigma]:=\prod_{\alpha}\delta[\sigma_\alpha]$ denotes the analogous product of delta functionals of all gauge fixing conditions. We see from this expression that interchanging the role of a gauge-fixing condition and a constraint has {\bf no} effect on the formal path integral as long as the set of constraints remains first class. We can thus express the Shape Dynamics path integral as a path integral where all ADM constraints and all conformal constraints appear as gauge-fixing conditions and constraints, because it does not matter in which place they appear.

Let us now for simplicity assume that we Abelianized all constraints\footnote{This is locally (in phase space) always possible, since the constrainst are regular.}, so we can follow the canonical DeWitt-Faddeev-Popov procedure and construct an action such that the path integral representation implements the constraints. These formal manipulations yield the Shape Dynamics path integral 
\begin{equation}
Z=\int Dg_{ab}\,D\pi^{ab}\,D\eta_\alpha\,DP^\beta \exp\left(i\int dt(\,d^3x\, (\pi^{ab}\dot g_{ab}+P^\alpha\dot \eta_\alpha)-\sigma_\alpha\chi^\alpha-\eta_\alpha \{\chi^\alpha,\sigma_\beta\}P^\beta)\right).
\end{equation}
The bare Hamiltonian $H=\sigma_\alpha\chi^\alpha+\eta_\alpha \{\chi^\alpha,\sigma_\beta\}P^\beta=\{\Omega,\bar \Omega\}$ in the gauge-fixed canonical path integral has now two BRST-invariances, which are generated by the Abelian BRST charges
\begin{equation}
\Omega=\eta_\alpha \chi^\alpha\,\textrm{ and }\,\bar\Omega=P^\alpha\sigma_\alpha,
\end{equation}
where $\Omega$ has ghost number $+1$ and $\bar \Omega$ has ghost number $-1$, if the set of gauge fixing conditions is itself Abelian. This double BRST-invariance is possible {\it only} if the gauge-fixing surface is first class and due to the fact that the gauge-fixed Hamiltonian is the Poisson-bracket of a BRST-charge and an opposite-BRST-charge\footnote{We say opposite and not \lq\lq{}anti\rq\rq{} because we are not dealing with what is usually referred to as anti-BRST invariance. Usual anti-BRST-invariance requires that the two charges strongly commute and is due to a redundancy in constraints (primary and secondary) and a reducibility expressed in terms of Lagrange multipliers.}, which do not commute.

This simple application of the DeWitt-Faddeev-Popov procedure required that we Abelianized the constraints before formulating the path integral.  This Abelianization however introduces a significant amount nonlocality, which we want to avoid, since having a local bare action is extremely valuable and the reason for using the DeWitt-Faddeev-Popov procedure in the first place. We thus use the DeWitt-Faddeev-Popov procedure simply as a motivating example.

The way the original BRST invariance was observed, was by directly constructing a transformation of the extended phase space variables that did not change the path integral. We should note however, that had we  introduced non-abelian constraints to the Fadeev-Popov procedure, defining  a  nilpotent BRST transformation for the extended variables becomes impossible. Since our main aim in this paper is identifying local symmetry transformations through the BRST mechanism, and the ADM constraints are non-abelian (and thus would require non-localities were we to abelianize them), we have to introduce the BRST construction. This is what we turn to now.  Later we will use the full canonical BRST formalism for constructing a BRST-gauge-fixed Hamiltonian of the form $H=\{\Omega,\bar \Omega\}$, where $\Omega$ is a nilpotent BRST charge of ghost number $+1$ and $\bar \Omega$ is a nilpotnent BRST charge of ghost number $-1$, which is then automatically invariant under two sets of BRST-transformations. Our plan is thus to construct an extended version of Shape Dynamics, which is a first class system of constraints that gauge-fixes ADM. After giving this construction, we will return to symmetry doubling of usual Shape Dynamics.

\subsection{General BRST}

The fermionic extension of phase space required by BRST formally resembles the introduction of connection forms in usual gauge-theories in order to make them locally covariant. This extended phase space is equipped with a grading, (which is analogous to the degree of exterior algebras)  given by the ghost number, and a nilpotent graded differential, given by the BRST transform, which is analogous to a vertical derivative.\footnote{Higher ghost numbers (higher degrees of the exterior algebra) come in the construction because one must reconcile a projection onto the constraint surface (the Koszul differential), and the gauge orbit projection (the vertical derivative) into a single nilpotent operator, which provides the required cohomology. } This provides a resolution of the observable algebra, meaning that the observable algebra can be identified with the cohomology of the BRST differential at vanishing ghost number.

Let us now formulate symmetry trading in a generally covariant theory in the BRST formalism. We start with revisiting the BRST formalism for the (bosonic) system $(\Gamma,\{.,.\},\{\chi_\alpha\}_{\alpha \in \mathcal A})$ above. For this we adjoin to $\Gamma$ an extension by fermionic ghosts $\{\eta^\alpha\}_{\alpha \in \mathcal A}$ and canonically conjugate ghost momenta $\{P_\alpha\}_{\alpha \in \mathcal A}$, such that there is one ghost-antighost pair for each constraint.

What we want, and what is guaranteed by homological perturbation theory, is a nilpotent differential $s$ that incorporates both the Kozsul-Tate differential $\delta$ (associated to projection onto the constraint surface) and the longitudinal derivative $d_{\mbox{\tiny V}}$ (associated to projection by the gauge orbits). By incorporation, one means that it has to reproduce their action at ghost number zero. It is easy to see that it requires the symplectic realization of $s$ (i.e. its realization as an extended phase space function acting through the extended Poisson bracket) is just:
\be \Omega= \eta^\alpha\chi_\alpha+\order{\eta^2}
\ee

 The ghost-number one fermionic BRST generator $\Omega$ is then defined as
\begin{equation}
 \Omega=\sum_{n=0}^\infty \eta^{\alpha_1}...\eta^{\alpha_{n+1}} U^{(n) \beta_1 ... \beta_n}_{\alpha_{n+1}...\alpha_1} P_{\beta_1}...P_{\beta_n},
\end{equation}
where the zeroth order structure functions are $U^{(0)}_\alpha=\chi_\alpha$ and the higher order structure functions $U^{(n) \beta_1 ... \beta_n}_{\alpha_{n+1}...\alpha_1}$ are chosen such that $\Omega$ is nilpotent, i.e.  $\{\Omega,\Omega\}=0$, which implies relations order by order in ghost number. A sufficient solvability condition for the higher structure functions is that the set $\{\chi_\alpha\}_{\alpha \in \mathcal A}$ is first class. If a given system is abelian we have:
\be  \Omega= \eta^a\chi _a,
\ee
which can be readily checked to be nilpotent. If the constraints are first class and the structure functions are constants,  i.e.  the algebra of constraints is not ``soft", then the BRST charge is of \emph{rank one} and comes in the following form:
\be   \Omega= \eta^a\chi _a-\frac{1}{2}\eta^b\eta^aU_{ab}^c P_c
\ee
The \emph{rank} of a system can be identified with the order of ghost momenta required for constructing a nilpotent BRST charge. Although as we will see, ADM gravity has a soft first class constraint algebra, it is still of rank one, meaning that it does not possess terms with two or more ghost momenta. For more complicated systems like the relativistic membrane or the bosonic string, the construction has to proceed, order by order, until one finds a nilpotent function reproducing $\delta$ and $d_{\mbox{\tiny V}}$ at first order.

The gauge-fixed Hamiltonian is constructed by choosing a ghost number $-1$ fermion $\tilde\Psi=\tilde\sigma^\alpha P_\alpha+...$, where $\{\tilde\sigma^\alpha\}_{\alpha \in \mathcal A}$ is a set of proper gauge fixing conditions. 
Denoting the BRST invariant extension of the on-shell Hamiltonian (where all constraints are set to vanish) by $H_o$, the general gauge fixed BRST-Hamiltonian is written as
\begin{equation}
 H_{\tilde\Psi}=H_o+\eta^\alpha V_\alpha^\beta P_\beta+\{\Omega,\tilde\Psi\},
\end{equation}
where $\{H_o,\chi_\alpha\}=V_\alpha^\beta\chi_\beta$. The gauge fixing is attained by changing the BRST invariant extension of the Hamiltonian by a BRST-exact term. This changes the dynamics of ghosts and other non-BRST invariant functions, but maintains evolution of all BRST invariant functions. The crux of the BRST-formalism is that the gauge-fixed Hamiltonian $H_{\tilde\Psi}$ commutes strongly with the BRST generator $\Omega$. Although  gauge symmetry is completely encoded in the BRST transformation $s  := \{\Omega, . \}$, and we have fixed the gauge, the system retains a notion of gauge-invariance through BRST symmetry.

\subsection{Symmetry doubling in BRST}

Applying this to a generally covariant theory, i.e. a system with vanishing on-shell Hamiltonian $H_o=0$, we find that the gauge-fixed BRST-Hamiltonian takes the form
\begin{equation}
 H_{\tilde\Psi}=\{\Omega,\tilde\Psi\}.
\end{equation}
Now comes the rather simple central insight that makes symmetry doubling possible: since the set $\{\sigma^\alpha\}_{\alpha \in \mathcal A}$ is both a proper gauge fixing for $\chi_\alpha$ and a first class set of constraints, one can construct a nilpotent gauge-fixing $\Psi$ analogous to the construction of the BRST generator; the only difference is that ghosts and antighosts are swapped:
\begin{equation}
 \Psi=\sum_{n=0}^\infty P_{\alpha_1}...P_{\alpha_{n+1}} V_{(n) \beta_1 ... \beta_n}^{\alpha_{n+1}...\alpha_1} \eta^{\beta_1}...\eta^{\beta_n},
\end{equation}
where $V_{(0)}^{\alpha}=\sigma^\alpha$ and the first class property of the set $\{\sigma^\alpha\}_{\alpha \in \mathcal A}$ is sufficient for existence of the higher structure functions $V_{(n) \beta_1 ... \beta_n}^{\alpha_{n+1}...\alpha_1}$. The result for rank one theories is given simply by
\be \Psi=\sigma^\alpha P_\alpha-\frac{1}{2} P_b P_aC^{ab}_c \eta^c
\ee
Where we once again note that the BRST charge of the $C'$ system is obtained from the gauge fixing fermion by substituting ghosts by the ghost momenta, and vice-versa.

This means that the Hamiltonian is invariant under two BRST transformations
\begin{equation}
 \begin{array}{rcl}
   s_1 . &=& \{ \Omega , . \} \\
   s_2 . &=& \{ . , \Psi \} ,
 \end{array}
\end{equation}
which follows  directly from the super-Jacobi identity and nilpotency of both $\Omega$ and $\Psi$. It should be emphasized that one of the strengths of the BRST treatment is that such symmetries are not restricted to be on-shell, but take effect all over phase space. \footnote{As mentioned before, the geometrical analogy with gauge theories is quite strong. Just as in gauge theories one introduces a connection form with certain transformation properties to make local gauge transformation act in the same way as global gauge transformations (homogeneously), here one introduces ghosts (and ghost momenta) and new transformation properties of these, making the extended symmetry to be not just restricted to the constraint surface.} The dual structure of the gauge-fixing fermion with respect to the initial system, and the double invariance of the gauge-fixed Hamiltonian, allows us to see rather straightforwardly that in fact the two symmetries are acting in the exact same way on our dynamical system.  One is a mirror of the other, albeit acting in different ways on the dynamical variables, and no one symmetry can be seen as ``the gauge-fixing" and the other as the ``true symmetry". The  system possesses both symmetries, in an equal footing. 

Let us stop here for a second to take stock, what the necessary conditions for the symmetry doubling are:
\begin{enumerate}
\item That we initially have a local, generally covariant system (its Hamiltonian is pure constraint) of regular irreducible first class constraints. 
\item That we find a local gauge fixing of the above constraints, such that the gauge-fixing functions are themselves first class. 
\end{enumerate}  
It makes it easier that in our case the gauge-fixing fermion turns out to be of rank one as well, but in principle the framework works in the same way for general open and soft algebras. 
From the results of section \ref{sec:canonical} (e.g. see proposition \ref{prop:general_sym_trading} we can now straightforwardly see that whenever symmetry trading is possible, so is symmetry doubling. We thus state:
\begin{theo}
When the conditions given in proposition \ref{prop:general_sym_trading} are satisfied, both symmetry trading and the BRST symmetry doubling can be constructed. 
\end{theo}

\subsection{Observables}

Let us now take a detour from the main argument in this paper and consider observables: We can not prove the existence of a complete set of doubly BRST-invariant observables unless both sets of constraints can be simultaneously Abelianized (i.e. with one canonical transofrmation on the BRST-extended phase space). One can of course locally (in phase space) Abelianize both sets of constraints as we did previously. This Abelianization does however generically introduces nonlocalities in field theories that can not simultaneously undone by a canonical transformation on the BRST-phase space. So, generically one ends up with a system that has one nonlocal set of constraints, which is problematic.

In the simultaneously Abelianizable case we can proceed as follows: Observables are by definition smooth functions on the intersection $\mathcal C \cap \mathcal C^\prime$, i.e. the observables are elements of $C^\infty(\mathcal C \cap \mathcal C^\prime)$. Since both $\Gamma$ and $\mathcal C \cap \mathcal C^\prime$ are assumed to be symplectic and both sets of constraints are regular, we are locally able to choose almost-Darboux coordinates $(q_\mu,p^\nu;\chi_\alpha,\sigma^\beta)$ where $(q_\mu,p^\nu)$ coordinatize $\mathcal C \cap \mathcal C^\prime$ and Poisson commute with the $(\chi_\alpha,\sigma^\beta)$. Observables can now be identified with phase space functions $f(q,p)$ that depend on $(q_\mu,p^\nu)$ only, but are independent of $(\chi_\alpha,\sigma^\beta)$. The assumption of simultaneous Abelianizability is that there is an invertible operator $M^\alpha_\beta$ such that both $\{\tilde \chi_\beta=M^\alpha_\beta\chi_\alpha\}_{\beta \in \mathcal A}$ and $\{\tilde \sigma^\alpha=\sigma^\alpha M^\alpha_\beta\}_{\alpha \in \mathcal A}$ are Abelian sets of constraints. The BRST generators then take the form
\begin{equation}
 \Omega=\eta^\alpha \tilde \chi_\alpha \textrm{ and } \Psi=\tilde \sigma^\alpha P_\alpha.
\end{equation}
It follows straightforwardly that any observable $f(q,p)$ commutes with both $\Omega$ and $\Psi$. We can now go back to the original system using a canonical transformation on extended phase space generated by by the exponential Poisson action of $C=\eta_\alpha \epsilon^\alpha_\beta P^\beta$, where $M$ is the exponential of $\epsilon$. This can not change the pure ghost-number zero part of $f$, since Poisson brackets with $C$ can not decrease pure ghost- or pure antighost- number. We thus find in the simultaneously Abelianizable case that for each observable $f_o(q,p)$ there exists an $f(q,p,\chi,\sigma,\eta,P)$ that commutes with both $\Omega$ and $\Psi$ and satisfies the strong equation $f|_{\eta=0=P}=f_o$. 

Let us now consider the case when the constraints can not be simultaneously Abelianized. In this case we can proceed as above, but upon Abelianization the BRST generators take the form $\Omega=\eta^\alpha \tilde \chi_\alpha + \mathcal O(\eta^2)$ and $\Psi=\tilde \sigma^\alpha P_\alpha.$ One can now use the standard method to construct and extension $f$ for a given observable $f_o(q,p)$, that commutes with $\Omega$, but now one has to solve
\begin{equation}
  \{f,\Psi\}+\{\{\Omega,Q\},\Psi\}=0
\end{equation}
for $Q$, which may not be solvable. It may thus in general only be possible to have a complete set of observables that commutes with one of the BRST-generators.

Let us now return to the main argument and note that the gauge-fixed action 
\begin{equation}
  S=\int dt\left(\dot{q}_I p^I + \dot \eta_\alpha P^\alpha - \{\Omega,\Psi\}\right)
\end{equation}
is simultaneously invariant under the pair of BRST transformations $s_1$ and $s_2$, irrespective of the question whether there is a complete set of observables that is simultaneously invariant under the two BRST transformations. We note that  {\it symmetry doubling} in a very concrete sense extends to the quantum regime,  because classical BRST symmetries lead to symmetries of the quantum effective action through the corresponding Zinn-Justin equations.

\section{Doubly General Relativity}\label{sec:DGR}

\subsection{Construction}

Now we construct a gravity theory with symmetry doubling. This will amount, in a sense which we will make clear in this section, that  we use the local conformal symmetry as a gauge-fixing for ADM gravity. The main technical device of this paper, is to form a gauge-fixing fermion for ADM from a BRST charge corresponding to the symmetries of what we here call the  conformal theory, by swapping ghosts for ghost momenta in that charge. Let us construct this step by step.

The BRST-charge for irreducible (first class\footnote{The BRST symmetries are only well-defined for first class systems.}) constraints, forming a rank one system  is given by $\Omega= \eta^a\chi _a-\frac{1}{2}\eta^b\eta^aU_{ab}^c P_c$
where summation includes integration in the case of continuous variables,  $P^c$ are the ghost momenta associated to the first class constraints and $U_{ab}^c$ are the structure functions  for the first class constraints $\{\chi_ a\} $. In this condensed abstract index notation, subscripts stand in for  both the continuous and discrete variables. Note that the symmetries of the ghost fields are compatible with the  symmetries of the structure functions, so that $\eta_1\eta_2U_{12}^c =\eta_2\eta_1U_{21}^c$, for two constraints enumerated by $1$ and $2$. 

The only non-zero  elements of the ADM constraint algebra matrix are:
\begin{eqnarray*} U_{S(x), S(y)}^{H_ a(z)}&=&g^{ac}\left(\delta(z,x)\delta(z,y)_{;c}-\delta(z,y)\delta(z,x)_{;c}\right)\\
  U_{S(x), H_a(y)}^{S(z)}&=&\delta(z,y)_{;a}\delta(z,x)\\
 U_ {H_a(x), H_b(y)}^{H_c(z)}&=&\delta^c_b\delta(z,x)\delta(z,y)_{;a}-\delta^c_a\delta(y,z)\delta(z,x)_{;b}
\end{eqnarray*} 

Although these structure functions do not let us assume from the onset that the ADM constraints form  a rank one system, it turns out that this is indeed the case. Thus we can write for the ADM BRST charge:
\begin{equation}\label{equ:BRST_ADM}
 \Omega_{\mbox{\tiny ADM}}=\int d^3x \left(\eta S+\eta^ag_{ac}\pi^{cd}_{;d}+\eta^b\eta^a_{,b}P_a+\frac{1}{2}\eta^a\eta_{,a}P+\eta\eta_{,c}P_{b}g^{bc}\right)
\end{equation}
Here the ghosts associated with the scalar constraints are $\eta(x)$, while the ones associated with the momentum constraints are $\eta^a(x)$. The ghost momenta are denoted analogously. An explicit calculation shows that this definition yields $ \{\Omega_{\mbox{\tiny ADM}},  \Omega_{\mbox{\tiny ADM}}\}=0$. 

 The BRST-charge for the conformal theory \footnote{Note that the conformal theory is clearly not Shape Dynamics, as it possesses full Weyl invariance, not volume-preserving ones, and no true Hamiltonian, as does SD. } is
$$ \Omega= \eta^a\sigma _a-\frac{1}{2}\eta^b\eta^aC_{ab}^c P_c
$$
And the non-zero structure function is:
$$ C_{\pi(x), F^b(y)}^{F^ a(z)}=\delta(x,y)\delta(y,z)\delta_a^b
$$
In this case there is no dependence of the structure functions on the phase space variables. Since switching the role of the ghosts and ghost momenta does not change the BRST construction (it does change the density weights however), we find the preferred gauge-fixing fermion to be:
\begin{equation}\label{equ:GFBRST_C}   
  \bar\Omega_{\mbox{ \tiny C}}=\int d^3 x\left (P\frac{\pi}{\sqrt g} +P_aF^a+\frac{1}{2}\frac{P}{\sqrt g}P_a\eta^a\right) 
\end{equation}
Note that in this way the tensorial structure of the  ghost and ghost momenta variables in the  BRST charge \eqref{equ:BRST_ADM} and in the gauge-fixing fermion \eqref{equ:GFBRST_C} felicitously match, and thus we can identify the ghost moment appearing in \eqref{equ:GFBRST_C} as the canonically conjugate pair of the ghosts appearing in \eqref{equ:BRST_ADM}, and vice-versa, without the need to use the metric to contract indices.

Now using the fact that the symplectic structure imposed on the extended phase space is taken to satisfy the canonical commutation relation between ghosts and ghost momenta, we have the following expression for the Hamiltonian of Doubly General Relativity (DGR):
\begin{subequations}\label{equ:gf_BRST_Ham}
\begin{eqnarray}
H_{\mbox{\tiny{DGR}}}:=\{ \Omega_{\mbox{\tiny ADM}},   \bar\Omega_{\mbox{ \tiny C}}\}=
\label{etaP} \int d^3 x\Big((S+2\eta_{,c}P_bg^{bc}-\frac{1}{2}(\eta^aP)_{;a})\frac{1}{\sqrt g}(\pi-\frac{1}{2}P_a\eta^a)\\
\label{eta^aP_a}+(\pi_{a\phantom{b};b}^{\phantom{a}b}+\eta^b_{\phantom{b};a}P_b+(\eta^bP_a)_{;b}+\frac{1}{2}\eta_{,a}P)(F^a-\frac{1}{2\sqrt g}P\eta^a)+(\eta^b\eta^a_{,b}+\eta\eta_{,c}g^{ac})(\frac{1}{2\sqrt g}PP_a)\\
\label{SpiSF}-P(2(\nabla^2\eta-\eta R)-\frac{3}{2\sqrt g}\eta S)+( \hat\nabla_aP_b(\pi^{ab}-\frac{1}{3}\pi g^{ab})+\frac{1}{3}P_a(\Delta^b_{cb}\pi^{ac}+\frac{1}{2}\hat\nabla_bg^{ab}))\eta\\
\label{HF}-(g^{ab}P_c+\frac 1 3 P_kg^{ak}\delta^b_c)(\hat \nabla_a\hat \nabla_b \eta^c +\hat {R^c}_{bda}\eta^d)+P_a\mathcal{L}_{\vec{\mathbf\eta}}F^a\\
\label{HpiHgSgGammapigpi}+\frac{P}{\sqrt g}\pi^{ab}\mathcal{L}_{\vec{\mathbf\eta}}g_{ab}+\frac{\eta^a_{\phantom{a};a}}{\sqrt g}P(\pi+P_a\eta^a)-\frac{1}{4}\eta P\eta^aP_ a\frac{\pi}{\sqrt g}+\eta P^a\eta_{,a}P\Big)
\end{eqnarray}
\end{subequations}
We enumerate the lines in the above equation according to the following order: line \eqref{etaP} denotes the terms coming from the bracket $\{\eta, P\}$, \eqref{eta^aP_a} from $\{\eta^a, P_a\}$, \eqref{SpiSF} contains the terms involving $\{S,\pi\}$ and $\{S,F^a\}$, \eqref{HF} refers to $\{H_ a, F^b\}$, and the terms from the last line \eqref{HpiHgSgGammapigpi}  refer to the terms containing the brackets $\{H^a, \pi\}, \{H^a, \frac{1}{\sqrt g}\},  \{S, \frac{1}{\sqrt g}\} , \{g^{ab}, \pi\}$, respectively. 
A number of simplifications are possible, but we refrain from writing down the resulting, similarly complex, equation since its specific form  is not more illuminating than the present one. 

The above equation defining the ADM gauge fixed BRST extended Hamiltonian, $\{ \Omega_{\mbox{\tiny ADM}},   \bar\Omega_{\mbox{ \tiny C}}\}$  contains two sets of invariances, one referring to the full set of ADM constraints (which are taken to represent the group of full 4-dimensional diffeomorphisms, although this interpretation holds only on-shell), and the other referring to full dilatation transformations and another symmetry, generated by $F^a$, which we do not attempt to interpret here. 

Let us write down how the two BRST charges \eqref{equ:BRST_ADM} and \eqref{equ:GFBRST_C} act on the basic variables $g_{ab},\pi^{ab}, \eta, P, \eta^a, P_a$: 
\begin{eqnarray*}
\{ \Omega_{\mbox{\tiny ADM}}, g_{ab}\}&=&-\frac{\eta}{\sqrt g}(\pi_{ab}-\frac{1}{2}\pi g_{ab})-\mathcal{L}_{\vec{\mathbf\eta}}g_{ab}
\\
\{ \Omega_{\mbox{\tiny ADM}}, \pi^{ef}\}&=&
\left(-\frac{1}{2\sqrt g} g^{ef}G_{abcd}\pi^{ab}\pi^{cd}+\frac{2}{\sqrt g}(\pi^{eb}g_{bd}\pi^{fd}-\frac{\pi^{ef}\pi}{2})\right)\eta-\Big(\frac{1}{2}\sqrt g g^{ef} R\eta\nonumber\\
&+&\sqrt g(-R^{ef}-g^{ef}\nabla^2\eta+
\eta^{;ef} )\Big)+\mathcal{L}_{\vec{\mathbf\eta}}\pi^{ab}-\eta\eta_{,a}P_{b}g^{be}g^{af}\\
 \{ \Omega_{\mbox{\tiny ADM}}, \eta\}&=&-\frac{1}{2}\eta^a\eta_{,a}\\
 \{ \Omega_{\mbox{\tiny ADM}}, P\}&=&S- \frac{1}{2}(\eta^aP)_{,a}+2\eta_{,c}P_b g^{bc}\\
\{ \Omega_{\mbox{\tiny ADM}},\eta^a\}&=&-\eta^b\eta^a_{\phantom{a},b}-\eta\eta_{,c}g^{ac}\\
\{ \Omega_{\mbox{\tiny ADM}},P_a\}&=&-\pi_{a\phantom{c};c}^{\phantom{a}c}+(\eta^bP_a)_{,b}+\eta^b_{\phantom{a},a}P_b
\end{eqnarray*}
\begin{eqnarray*}
\{ \bar\Omega_{\mbox{\tiny C}}, g_{ab}\}&=&\frac{P}{\sqrt g}g_{ab} \\
\{ \bar\Omega_{\mbox{\tiny C}}, \pi^{ef}\}&=&\frac{P}{\sqrt g}(\pi^{ab}-\frac{1}{ 2} g^{ab}(\pi+\frac{1}{2} P_c\eta^c))-(P_k\Delta^k_{ab}-\frac{1}{3}P_b\Delta^k_{ak})g^{ea}g^{fb}-g^{fb}g^{ka}P_{a;b}+\frac{1}{3}g^{ef}g^{kd}P_{k;d}\\
\{ \bar\Omega_{\mbox{\tiny C}}, \eta\}&=&-\frac{1}{\sqrt g}(\pi+\frac{1}{2} P_c\eta^c)\\
\{ \bar\Omega_{\mbox{\tiny C}}, P\}&=&0\\
\{ \bar\Omega_{\mbox{\tiny C}}, \eta^a\}&=&-F^a+\frac{P}{2\sqrt g}\eta^a\\
\{ \bar\Omega_{\mbox{\tiny C}}, P_a\}&=&\frac{P}{2\sqrt g}P_a
\end{eqnarray*}

\subsection{Consequences}

Let us stop for a moment to take stock of some of the implications of the doubly invariant gauge-fixed Hamiltonian constructed above. 

\subsubsection*{The Hamiltonian at ghost number zero}

The Hamiltonian at ghost number zero, straightforwardly read off from \eqref{equ:gf_BRST_Ham}, is given by:
\be \label{equ:BRST_Ham_gh(0)}
H_{\mbox{\tiny{DGR}}}=S(\pi-\lambda\sqrt g)+H_a(F^a)+\order{\eta}
\ee
Thus it is interesting to note that this is \emph{neither} the frozen lapse Hamiltonian ($N=0$) nor the CMC Hamiltonian $(H(N_{\mbox{\tiny{CMC}}})$, but it is still a generator of dynamics, with third order dependence in time-derivatives of the metric (cubic momenta appear). 

Although third order time derivatives are usually associated to instabilities, in our case this is merely a gauge-fixing of the physical Hamiltonian, and thus has to be well-defined for gauge-invariant observables. One can see that the evolution of  observables indeed obeys the correct properties:
if we act with it on some phase space function $A$, we get for the first two terms: $\{H_{\mbox{\tiny{DGR}}},A\}=\{S,A\}(\pi-\lambda\sqrt g)+S\{\pi-(\lambda\sqrt g),A\}+\dots$. Now it is easy to see that if A is an ADM (resp. E-SD) observable, the only non-weakly vanishing term is proportional to $S$ (resp. $\pi-\lambda\sqrt g$), which weakly vanishes on the ADM (resp. E-SD) constraint surface. This is only a heuristic way of seeing the more general fact that 
$$ \{H_{\mbox{\tiny{DGR}}},A\}=\{\{ \Omega_{\mbox{\tiny ADM}},   \bar\Omega_{\mbox{ \tiny C}}\},A\}= \{\Omega_{\mbox{\tiny ADM}}, \{  \bar\Omega_{\mbox{ \tiny C}},A\}\}
$$ 
which is thus ADM-BRST exact.

\subsubsection*{Anti-BRST}

In usual BRST analysis one can introduce a redundant symmetry into the system, analogously to the usual Kretschmanization, called an anti-BRST symmetry. Let us call this extra BRST symmetry $\Omega_{\mbox{\tiny{Anti-ADM}}}$. Assuming that the initial constraints are abelian for simplicity, when one ``doubles" the amount of identical constraints, $\chi_1=\chi_2=\chi$, thereby obtaining a reducible set of constraints with the extra reducibility condition $\chi_1-\chi_2=0$. In this case the BRST operator $\Omega_{\mbox{\tiny{total}}}=\eta_1\chi^1+\eta_2 \chi^2+\lambda(P_1-P_2)$ obeys a bi-degree expansion \cite{AntiBRST} that yields two commuting BRST generators:
$$\Omega=\eta_1\chi^1+\lambda P_1~, ~\mbox{and}~\Omega_{\mbox{\tiny{Anti}}}=\eta_2\chi^2-\lambda P_2
$$
But, one of the defining properties of the anti-BRST charge is that it commutes with the BRST charge. Namely, the algebra for anit-BRST is the following:
$$ \{ \Omega,   \Omega_{\mbox{\tiny{Anti}}}\}=\{ \Omega,  \Omega  \}=\{\Omega_{\mbox{\tiny{Anti}}},   \Omega_{\mbox{\tiny{Anti}}}\}=0
$$
Whereas the algebra of the DGR charges follow a supersymmetric algebra:
$$ \{ \Omega_{\mbox{\tiny ADM}},   \bar\Omega_{\mbox{\tiny{C}}}\}=H_{\mbox{\tiny{DGR}}}~,~\{ \Omega_{\mbox{\tiny ADM}},  \Omega_{\mbox{\tiny ADM}}  \}=\{\bar\Omega_{\mbox{\tiny{C}}},   \bar\Omega_{\mbox{\tiny{C}}}\}=0
$$

\subsubsection{SUSY}

There is a rather straightforward argument why Shape Dynamics should not suffer from Weyl anomalies, even though it does possess spatial Weyl symmetry. The argument is that the Linking Theory for ADM is merely a Kretschmannization (the introduction of an additional degree of freedom that is declared to be pure gauge), and as the metric sector of ADM (unlike the internal Lorentz sector) does not possess anomalies, the Kretschmannized theory cannot be anomalous, because anomalies are of topological nature,\footnote{Topological here of course does not refer to the topology of the underlying space, but rather to the infinite dimensional space of sections over configuration space. } it would be interesting to see this mechanism arises from our ``supersymmetric algebra" akin to the usual anomaly cancellation from supersymmetry mechanisms in quantum field theory. A hint for this is the formulation of the classical partition function (i.e. the kernel that is only supported on classical boundary data) in the path integral language, which turns out to be be symmetry doubling precisely in the sense described in this paper (compare e.g. \cite{Reuter}).

\subsubsection*{Observation: relation with master constraint}

It is interesting to note that the present work bears some relationship with the master constraint program, it generalizes it and makes it more transparent in the BRST framework. 

The master constraint program is motivated very differently, but has
an interpretation in terms of a degenerate case of the symmetry
doubling framework presented in this paper. For simplicity we restrict
the presentation to the Abelian case and assume that all constraints
$\chi^\alpha$ strongly commute. The master constraint for this system
is
\begin{equation}
 M=\chi^\alpha \delta_{\alpha\beta} \chi^\beta
\end{equation}
Weak observables $O$ are detected by the condition
$\{\{M,O\},O\}|_{M=0}=0$, which by use of the Jacobi identity is
equivalent to
\begin{equation}
 0=\{\{M,O\},O\}|_{M=0}=2\{\chi^\alpha,O\}\delta_{\alpha\beta}\{\chi^\beta,O\}|_{\chi\equiv
0},
\end{equation}
which is equivalent to the weak observable condition
$0=\{\chi^\alpha,O\}|_{\chi\equiv 0}$.

Let us now compare this to the degenerate case of the construction
described in this paper where we use the $\chi^\alpha$ for the
construction of $\Omega$ and $\bar \Omega$, which read
\begin{equation}
 \Omega=\eta_\alpha \chi^\alpha\,\textrm{ and }\,\bar \Omega=P^\alpha
\delta_{\alpha\beta} \chi^\beta,
\end{equation}
so the doubly invariant Hamiltonian is precisely the master
constraint. The difference with the master constraint approach is that
BRST-observables obey the linear relation
$\{\Omega,O_{BRST}\}=0$ rather than the nonlinear relation above. One can easily see that the  master constraint is a degenerate case of the Symmetry Doubling scenario (it in fact is not doubling the symmetries) because after the action of its Hamiltonian (equiv. Hamiltonian at ghost number zero in BRST), only terms proportional to the original scalar constraint will be weakly zero  (equiv. BRST-closed). 

\subsection*{True Shape Dynamics}

In the previous sections we constructed Doubly General Relativity with complete symmetry trading, i.e. we also traded the spatial diffeomorphisms of ADM. The construction of a doubly invariant Hamiltonian does however not require complete symmetry trading. Moreover, from the relational first principles that motivated Shape Dynamics, it is conceptually more pleasing to retain the spatial diffeomorphisms also on the Shape Dynamics side. We thus use the following constraints to construct the dual BRST generator $\bar \Omega$: the conformal generators $D(x)=\frac{\pi(x)}{\sqrt{g}(x)}$ and the spatial diffeomorphism generators $H_a(x)=\frac{-2g_{ab}(x)\pi^{bc}_{;c}(x)}{\sqrt{g}(x)}$. 

This system seems to have many interesting features and will be investigated in a future paper.

\section{Effective Field Theory}\label{sec:EFT}

The purpose of this section is to combine standard effective field theory reasoning with a semiclassical expansion to derive that the effective action should indeed have the BRST-invariances of both ADM and E-SD \emph{in a semiclassical regime}.

Let us briefly review effective field theories. A good starting point is the functional renormalization group equation, which tells one how the effective average action $\Gamma_k[\phi]$ changes with the coarse graining scale $k$. The coarse graining scale is  implemented by inserting the IR-supression term $\frac 1 2 \langle \phi, R_k \phi\rangle$ \footnote{The infra-red (IR) suppression term gives a mass of order $k^2$ to modes below scale $k$ while vanishing for modes above $k$, so it vanishes for $k=0$}  into a formal path integral definition for the partition function $e^{W_k[j]}=\int d\mu[\phi]\exp(-\frac 1 2 \langle \phi,R_k \phi \rangle + \langle \phi,j\rangle)$, where $d\mu[\phi]$ denotes the full path integral measure which also contains the exponential of the bare action. The effective average action, defined as $\Gamma_k[\phi]=\sup_j\left(W_k[j]-\langle \phi,j\rangle\right)-\frac 1 2 \langle \phi,R_k\phi\rangle$ satisfies the flow equation:
\begin{equation}
 k\partial_k \Gamma_k[\phi] = \frac 1 2 \textrm{STr}\left(\frac{k \partial_k R_k}{\Gamma^{(2)}_k[\phi]+R_k}\right).
\end{equation}
We denote the multiplet containing all fields by $\phi$, the field dependent two-point function by $\Gamma^{(2)}$ and the supertrace by $STr$. The usual effective action $\Gamma[\phi]=\Gamma_{k=0}[\phi]$ is attained in the limit of vanishing IR-suppression. It is important to notice that this equation does not require a bare action as input. It can thus also be used in situations where no fundamental theory (i.e. no fundamental measure $d\mu[\phi]$) is known to study the flow of the effective action in the IR. All that is required for the flow equation to be applicable is that the physical system under investigation can be well approximated as a local field theory with field content $\phi$ below some scale $\Lambda$. This is the first ingredient for the definition of a theory space: One requires the effective average action to be well approximated by a local\footnote{The effective action is of course a very complicated nonlocal functional, but for practical purposes in which physics is probed at a single scale only, it can be very well approximated as a properly renormalized local action.} functional of the form
\begin{equation}\label{equ:EAAexpansion}
 \Gamma_k[\phi]=\sum_i g_i(k) O_i[\phi],
\end{equation}
where the $O_i[\phi]$ are elements of a basis for the expansion of local functionals.

A very important feature, that makes the use of flow equations for the study of effective field theories very valuable is universality associated with attractors (in particular fixed points) of the renormalization group flow. This is due to the fact that the flow often has only a small number of relevant and marginal directions emanating from a fixed point, which span the critical surface associated with a fixed point. This critical surface is IR attractive, so independently of where a more fundamental theory predicts the initial condition for $\Gamma_{k=\Lambda}[\phi]$ to be, this initial condition will be attracted by the critical surface and approach it in the limit $k\to 0$. One can thus predict the form of the effective action $\Gamma[\phi]$ up to the small number of parameters that is given by the coordinates of the critical surface. 

Another practically very important feature of the renormalization group flow is that it is compatible with Ward identities. Let $d\mu[\phi]$ denote the full path integral measure and suppose that $d\mu[\phi]$ is invariant under the infinitesimal field transformation $T:\phi \mapsto \phi + \epsilon s\phi$, where $\epsilon$ is infinitesimal, i.e. $d\mu[\phi]=d\mu[T\phi]$. Inserting the invariance into the path integral definition of the partition function gives the Ward identity
\begin{equation}
 0=\mathcal W_k\triangleright \Gamma_k:=\int d\mu[\phi]s(-\frac 1 2 \langle \phi,R_k \phi \rangle + \langle \phi,j\rangle)\exp\left(-\frac 1 2 \langle \phi,R_k \phi \rangle + \langle \phi,j\rangle\right).
\end{equation}
Repeating the same calculation with $k\partial_k e^{W_k[\phi]}$ yields that the Ward-identity is preserved under the flow
\begin{equation}
 k\partial_k \mathcal W_k\triangleright \Gamma_k=0.
\end{equation} 
This tells us that the flow stays on the (evolving) surface $\mathcal W_k=0$ if the initial condition is on this surface\footnote{Notice that the operator $\mathcal W_k$ that acts on the functional $\Gamma_k$ in general evolves with $k$!}. Moreover, one finds that the standard Ward identity is attained in the limit $k\to 0$, because the IR-suppression term vanishes for $k=0$. One can thus consistently impose gauge symmetries in an effective field theory framework and put symmetry requirements into the definition of a theory space and require that $\mathcal W_k\triangleright \Gamma_k=0$ holds.

A third practically important point is dimensional analysis: This is due to the fact that one studies the flow of dimensionless couplings (i.e. $\tilde g_i=k^{-dim(g_i)} g_i$) to study universality. Then $k\partial_k \tilde g_i = -dim(g_i) \tilde g_i + more$ where the anomalous dimension indicated by the $more$-term is small in weakly coupled systems. This means that if we have physical input telling us that the system is weakly coupled in the IR then we can approximate the renormalization group flow with the dimension of the respective operators. The critical surface is then spanned by power counting relevant and marginal coupling constants.

\subsection{Standard Reasoning}\label{sec:effective}

We have seen from the study of the renormalization group that (1) locality, (2) field content, (3) symmetry and (4) dimensional analysis can often be used to make predictions about the effective action in the IR even if no fundamental theory is available. It is straightforward to use dimensional analysis as an ordering principle and implement the field content and locality in the theory space. For this we expand the effective action $\Gamma_k$ as in equation (\ref{equ:EAAexpansion}) where one may truncate the sum to range only over low dimensional operators $O_i$. The implementation of symmetry requires a bit more explanation.

For this we note that gauge symmetries of a classical action $S$ are expressed as BRST-symmetries of the gauge-fixed bare action $S_{BRST}$, which appears in the exponent in the path integral. This of course requires the introduction of ghosts into the theory space and puts the additional restriction that each term in the expansion of the effective action has vanishing ghost number. The BRST-transformations are however in general not linear in the fields, which renders the Ward-identities practically useless. To gain linearity one usually proceeds to the Zinn-Justin equation and adds anti-fields $\phi^\dagger$ as sources to the BRST-transforms of the fields $\phi$, where the multiplet $\phi$ includes the ghosts. The extended classical action $S_{BRST}+(s\phi)^A\phi^\dagger_A$, where $s$ denotes the BRST differential, satisfies the classical master equation $(S,S)=0$, where $(.,.)$ denotes the Batalin-Vilkovisky antibracket. The classical master equation encodes the classical gauge symmetry of the bare action. In the absence of gauge anomalies, i.e. whenever there is a proper regularization and solution to the quantum master equation $\frac 1 2 (S,S)=i \hbar \Delta S$, where $\Delta$ denotes the Schwinger-Dyson operator, one can impose the gauge symmetry in the form of a quantum master equation. This is to say that we assume that there is a nilpotent transformation $s$ such that the infinitesimal transformation $\phi \to \phi+\epsilon s\phi$ leaves the measure $d\mu[\phi]$ invariant.

Having $s,d\mu[\phi]$ we can derive the Zinn-Justin equation for the effective action $\Gamma[\phi,\phi^\dagger]$, which is obtained as the equation satisfied by the effective average action in the limit $k=0$. For this we consider the BRST variation $\phi \mapsto \phi + \epsilon s\phi$ of the partition function
\begin{equation}
  \begin{array}{rcl}
    e^{W[j,\phi^\dagger]}&=&\int d\mu[\phi] \exp\left(-\frac 1 2 \langle \phi, R_k\phi\rangle+\langle s \phi,\phi^\dagger\rangle+\langle \phi, j\rangle\right)\\
                         &=&\int d\mu[\phi] \left(1+\epsilon(-\frac 1 2 s\langle \phi,R_k\phi\rangle+\langle s\phi,j\rangle) \right)\exp\left(-\frac 1 2 \langle \phi, r_k\phi\rangle+\langle s \phi,\phi^\dagger\rangle+\langle \phi, j\rangle\right)\\
                         &=:&e^{W[j,\phi^\dagger]}+\epsilon \mathcal W_k,
  \end{array}
\end{equation}
where in the middle line we insert the variation, use nilpotency of the BRST variation and invariance of the measure once again.
This shows that  $\mathcal W_k=0$. At $k=0$, where the regulator terms vanish, we find
\begin{equation}
 \langle \frac{\delta_R W}{\delta \phi^\dagger},j \rangle=\mathcal W_{k=0}=0.
\end{equation}
Using the $\Gamma[\phi,\phi^\dagger]:=\sup_j\left(W[j]-\langle\phi,j\rangle\right)$ and inserting the extremizing source $j_*[\phi,\phi^\dagger]$ we can write $\mathcal W_{k=0}$ in terms of the effective action using $\frac{\delta_R \Gamma_{k}}{\delta \phi^\dagger}=\left.\frac{\delta W_k}{\delta\phi^\dagger}\right|_{j=j_*}$
\begin{equation}
 \langle \frac{\delta_R \Gamma_{k=0}}{\delta \phi^\dagger},\frac{\delta_L \Gamma_{k=0}}{\delta \phi}\rangle=\mathcal W_{k=0}=0,
\end{equation}
which is known as the Zinn-Justin equation. Moreover, using the definition of the antibracket, we find that the Zinn-Justin equation can be expressed as a classical master equation for the effective action
\begin{equation}
 \frac 1 2 (\Gamma,\Gamma)=0.
\end{equation}
This means that by passing to the antifield formalism, we can impose classical antifield symmetry on the effective action $\Gamma$. We thus \emph{define the theory space for the effective field theory as the space of local functionals on the antifield extension of the field content that satisfy the classical master equation}. 

The Zinn-Justin equation does however not imply that the effective action is linear in the antifields, which is necessary for it to implement \emph{classical} BRST-invariance. To conclude this, we need additional input about $\Gamma$. The most practical one is to assume a semiclassical regime, where expectation values of products of field operators are to lowest order in an $\hbar$-expansion given by products of fields. If we restrict ourselves to the lowest order in $\hbar$, we also avoid the issue of quantum  anomalies, because these can only appear at higher orders in $\hbar$. This can be read of from relative $\hbar$ in the quantum master equation. Thus, assuming a semicalssical regime and \emph{neglecting all higher orders of $\hbar$, we find that the effective action is BRST-invariant} because
\begin{equation}
  0=\langle s\phi_A\rangle \frac{\delta_L\Gamma}{\delta \phi_A}=s \Gamma +\mathcal O(\hbar).
\end{equation}
 
\subsection{New Definition of Effective Field Theories}

Let us now assume that we have classical field theory with a local BRST-gauge-fixed Hamiltonian $H$ that is invariant under under two BRST transformations $s_1 H=\{\Omega_1,H\}=0$ and $s_2 H=\{\Omega_2,H\}=0$. This is the situation we observe in classical symmetry doubling and it implies that the canonical action is invariant under two BRST transofrmations. We thus consider the canonical path integral
\begin{equation}
Z=\int Dg_{ab}\,D\pi^{ab}\,D\eta_\alpha\,DP^\beta \exp\left(i\int dt(\,d^3x\, \pi^{ab}\dot g_{ab}+\dot\eta^\alpha P_\alpha+\{ \Omega_{\mbox{\tiny ADM}}, \bar\Omega_{\mbox{\tiny C}}\})\right).\end{equation}
Thus, from the arguments of the previous section, assuming that the measure of the path integral is invariant, BRST variation yields two effective Slavnov-Taylor identities: 
\be\label{equ:slavnov}
\mean {s_{\mbox{\tiny ADM}}\phi_A}\diby{_L\Gamma}{\phi_A}=0~, ~\mbox{and}~~\mean {s_{\mbox{\tiny C}}\phi_A}\diby{_L\Gamma}{\phi_A}=0
\ee
Again, the non-linearities in the BRST variations prevent us from expecting that the corresponding two Zinn-Justin equations will implement exact classical BRST invariances. But in a semi-classical approximation we do have that:
\be s_{\mbox{\tiny ADM}}\Gamma=\order{\hbar}~, ~\mbox{and}~~s_{\mbox{\tiny C}}\Gamma=\order{\hbar}
\ee

Thus, from these semi-classical considerations we define:
\begin{defi}
Given any local action functional of the fields $g_{ab}, \pi^{ab}, \eta^a, P_a$, BRST-closed under both $ s_{\mbox{\tiny ADM}}$ and $ s_{\mbox{\tiny C}}$, 
a gravity theory is defined as its ghost number zero part. 
\end{defi}

\section{Conclusions}\label{sec:conclusions}

The main result of this paper is that there is a local gravity action
that is equivalent to General Relativity, and that this action has a hidden BRST
symmetry due to the duality between General Relativity and
Shape Dynamics. Applying standard effective field theory reasoning to
this Doubly General Relativity theory refines the definition of a
gravity theory by adjoining to the spacetime gauge symmetries the additional
local spatial Weyl symmetry of Shape Dynamics. In one aspect, it also provides a
significant improvement over previous Shape Dynamics actions in the sense that
the canonical action and the Hamiltonian are here local.

Several independent technical results were necessary to establish the
main result:
\begin{enumerate}
 \item {\it General Linking Theory:} We provided an explicit construction of a linking theory that proves the equivalence of gauge theories that are described by two first class constraint surfaces that gauge-fix one another. This extends the previous result \cite{Gomes:2011zi} where a special construction of linking gauge theories through an implementation of best matching was given.
 \item {\it Relation of Observable Algebras:} Using the general linking theory, we constructed the explicit dictionary between the observable algebras of two equivalent gauge theories through first working out the observable algebra in the linking theory and then using the two partial phase space reductions that yield the two equivalent gauge theories.
 \item {\it Extended Shape Dynamics:} We constructed an extended version of Shape Dynamics, whose gauge symmetries are local spatial conformal transformations (dilations) and transformations of the metric momenta, whose Poisson brackets among one another and with the dilatations at each point in the Cauchy surface resemble the conformal algebra (dilatations and special conformal transformations).
 \item {\it True Shape Dynamics:} By choosing a gauge symmetry that also gauge fixes the 3-diffeomorphism constraints,  Extended Shape Dynamics gives full symmetry doubling, i.e. one has doubled the number of local first class constraints. It may however not be necessary to have full symmetry doubling and one might rather want to trade only the scalar ADM-constraints for local Weyl constraints. This system has a solid conceptual motivation and will be investigated in a future paper.
 \item {\it Symmetry Doubling:} We showed that whenever symmetry trading is possible then one can construct a BRST-gauge fixed action that is invariant under two BRST-transformations: One that encodes the original gauge symmetry and a second that encodes the gauge symmetry of the equivalent dual gauge theory.
 \item {\it Relation of BRST Observables:} In the case where the two dual constraint systems can be simultaneously Abelianized, there is a complete preferred set of representatives of BRST-observables that strongly commutes with both BRST-charges. In all other cases this does not hold, but one can still relate the observables of the two theories, by proceeding as in the Dirac formalism and using the linking theory to work out the relation.
 \item {\it Doubly General Relativity:} Using the BRST charges of Extended Shape Dynamics and ADM gravity, we constructed a BRST-gauge fixed Hamiltonian, and thus a BRST-gauge fixed canonical action, that is invariant under two BRST differentials; one encodes on-shell spacetime diffeomorphism symmetry, the other local spatial conformal symmetry.
\end{enumerate}
Some immediate consequences are:
\begin{enumerate}
 \item {\it Bulk-bulk duality:} The equivalence of ADM-gravity and Shape Dynamics holds, unlike more familiar bulk-boundary dualities, not only at a boundary of a spacetime, but the trajectories of ADM gravity in constant mean curvature gauge coincide with the trajectories of Shape Dynamics in a preferred gauge. The duality between the spacetime description and the conformal theory description is thus bulk-to-bulk. The doubly BRST-invariant action provides a direct way to see this duality.
 \item {\it SUSY algebra:} The algebra of the two BRST charges is, other than the more familiar anti-BRST algebra, a supersymmetry algebra. Both charges are nilpotent and their Poisson-bracket yields the gauge fixed Hamiltonian.
 \item {\it Gravity Theory Space:} Applying standard effective field theory reasoning to Doubly General Relativity yields a new scenario for effective gravity theories, at least in a semiclassical limit. If we take the field content to be the BRST-extension of ADM gravity, then we demand that the effective action is left invariant by the ADM-BRST transform, which encodes spacetime symmetries on shell, and by the Extended Shape Dynamics-BRST transform, which encodes local spatial conformal invariance. 
\end{enumerate}
Let us conclude this paper with an incomplete outlook, which consists
of several distinct directions:
\begin{enumerate}
 \item {\it Observable consequences:} The conventional theory space for effective ADM gravity theories requires the on-shell spacetime symmetry alone. The theory space of Doubly General Relativity requires in addition a notion of local spatial conformal invariance and is thus more restrictive. Both theory spaces contain the Einstein-Hilbert action, but one has to expect that the additional symmetry principle rules out many deformations of the Einstein-Hilbert action that are allowed by the ADM symmetry principles. It is thus formally possible to experimentally falsify Doubly General Relativity, by observing effective gravity dynamics (beyond Einstein-Hilbert) that is incompatible with the symmetry principles of Shape Dynamics.
 \item {\it Quantum Shape Dynamics as a definition for Quantum Gravity:} This paper focused on the construction of a doubly invariant action for gravity. The classical observable equivalence between Shape Dynamics and General Relativity \cite{Koslowski:2012uk} however implies that a quantization of Shape Dynamics can be interpreted as a quantization of General Relativity. This suggests to explore whether gauge-fixings of Shape Dynamics, that are not manifestly equivalent to ADM, may improve renormalizability.
 \item {\it Application to other systems:} The linking theory formalism is very generic; all that is needed is a gauge symmetry that is gauge-fixed by another gauge symmetry. The existence of a linking theory is sufficient for symmetry trading and symmetry doubling. It would be very interesting to apply these constructions to other systems of physical interest, in particular Yang-Mills theories.
\end{enumerate}

\section*{Acknowledgments}
HG would like to thank the Perimeter Institute for Theoretical Physics for hospitality. 
 HG was supported in part by the U.S.
Department of Energy under grant DE-FG02-91ER40674.
Research at the Perimeter Institute is supported in part by the Government of Canada through NSERC and by the Province of Ontario through MEDT.

\end{document}